\documentclass{emulateapj}

\newcommand{\beq}{\begin{equation}}
\newcommand{\eeq}{\end{equation}}
\newcommand{\TiO}{\rm [TiO]}

\newcommand{\teff}{$T_{\rm eff\ }$}

\def\earth{\hbox{$\oplus$}}
\def\arcmin{\hbox{$^\prime$}}
\def\arcsec{\hbox{$^{\prime\prime}$}}

\newcommand{\lsim}{\ \raise
-2.truept\hbox{\rlap{\hbox{$\sim$}}\raise5.truept\hbox{$<$}\ }}
\newcommand{\gsim}{\ \raise
-2.truept\hbox{\rlap{\hbox{$\sim$}}\raise5.truept\hbox{$>$}\ }}
\newcommand{\simsim}{\ \raise
-2.truept\hbox{\rlap{\hbox{$\sim$}}\raise5.truept\hbox{$\sim$}\ }}

\usepackage{palatcm}
\slugcomment{Accepted for publication
in the Astrophysical Journal Supplement Series.}
\shorttitle{Optical Photometry of the ONC}
\shortauthors{Da Rio et al.}

\begin{document}

\title{
\ A Multi-color Optical Survey of\ \ the Orion\ Nebula\ Cluster \\
\ Part I: the Catalog}

\author{N. Da Rio\altaffilmark{1}}
\affil{Max Planck Institut f\"{u}r Astronomie,  K\"{o}nigstuhl 17, D-69117 Heidelberg, Germany }
\altaffiltext{1}{Member of IMPRS for Astronomy \& Cosmic Physics
       at the University of Heidelberg}
\email{dario@mpia-hd.mpg.de}

\author{M. Robberto, D. R. Soderblom, N. Panagia\altaffilmark{2,3}}
\affil{Space Telescope Science Institute, 3700 San Martin Dr., Baltimore MD, 21218, USA\ }
\altaffiltext{2}{INAF-CT Osservatorio Astrofisico di Catania, Via S.Sofia 79, I-95123 Catania, Italy}
\altaffiltext{3}{Supernova Ltd, OYV \#131, Northsound Road, Virgin Gorda, British Virgin Islands}

\author{L. A. Hillenbrand}
\affil{California Institute of Technology, 1200 East California Boulervard, 91125 Pasadena, CA, USA}

\author{F. Palla}
\affil{INAF - Osservatorio Astrofisico di Arcetri, Largo E. Fermi, 5
I-50125 Firenze, Italy}

\and

\author{K. Stassun}
\affil{Vanderbilt Univ., Dept. of Physics \& Astronomy
6301 Stevenson Center Ln., Nashville, TN 37235, USA\ }




\begin{abstract}
We present $U$, $B$, $V$,  $I$ broad-band, 6200~\AA\ TiO medium-band and H$\alpha$ narrow-band photometry of the Orion Nebula Cluster (ONC)\ obtained with the WFI imager at the ESO/MPI 2.2 telescope at La Silla Observatory.  The nearly-simultaneous observations, cover the entire ONC in a field of about $34\times 34$ arcmin. They enable us to determine  stellar colors avoiding the additional scatter in the photometry induced by  stellar variability typical of pre-main sequence stars. We identify $2,612$ point-like sources in $I$ band, 58\%, 43\%
and 17\% of them detected also in $V$,\ $B$\ and $U$, respectively. 1040 sources are identified in H$\alpha$ band.
In this paper we present the observations, the calibration techniques adopted, and the resulting catalog. We show the derived CMD of the population and discuss the completeness of our photometry.
We define a spectro-photometric TiO\ index that takes into account the fluxes in V- I- and TiO-band. Comparing it with spectral types of ONC members in the literature, we find a correlation between the index and the spectral type valid for M-type stars, that is accurate to better than 1 spectral sub-class for M3-M6 types and better than 2 spectral subclasses for M0-M2 types.. This allows us to newly classify 217 stars.
In a similar way, we subtract from our H$\alpha$ photometry the photospheric continuum  at its wavelength, deriving  calibrated line excess for the full sample. This represents the largest H$\alpha$ star catalog obtained to date on the ONC.
This data set enables a full re-analysis of the properties of the Pre-Main Sequence population in the Orion Nebula Cluster to be presented, in an accompanying paper.
\end{abstract}



\section{Introduction}
The Orion Nebula Cluster (ONC)\ plays a fundamental role in our understanding of star and planet formation, serving, along with the Pleiades and Hyades, as a fundamental calibrator and prototype for young stellar clusters. As the nearest site where the entire Initial Mass Function (IMF), from $\simeq 25~M_\odot$ down to $10~M_{\rm Jup}$ can be studied   with minimal foreground and background contamination  \citep{jones-walker88,getman2005}, it has been extensively investigated both at visible \citep{herbig86,1994ApJ...421..517P,hillenbrand97,robberto04} and near-IR \citep{hillenbrand98disks,hillenbrand2000,lucasroche2000,luhman2000,slesnick04} and mid-IR \citep{robberto05,smith05}. In particular, the ONC\ provides a unique opportunity for studying a stellar cluster at visible wavelengths that is rich ($>2000$ members \citep{ali-depoy95}) and young ($\simeq 1~$Myr, \citet{hillenbrand97} [hereafter H97]).  In most cases clusters of this age are embedded within their parental environment \citep{ladalada2003}, but in the case of the ONC the neutral material has been removed by the expansion of the \ion{H}{2} region produced by the brightest cluster members, $\theta^1$ Ori C in particular.  By combining photometry and spectroscopy at visible wavelengths, it is possible to derive the extinction for each individual star, and therefore  its absolute luminosity, radius, and a model-dependent mass and age (H97).

The Hubble Space Telescope Treasury Program on the Orion Nebula Cluster
(HST\ GO-10246, P.I. M. Robberto) carried out multicolor visible photometry of the ONC with the highest possible spatial resolution and sensitivity, in order to obtain the most accurate estimates of stellar parameters. The Hubble observations have been complemented by ancillary ground based observations, necessary because of the limitations of the HST data in regard to source saturation and variability. Source saturation with HST was unavoidable given the long exposures (approximately 340~s) used to reach high signal-to-noise on the faintest ONC members. This time is long enough that point sources in Orion brighter than $I$~$\simeq 17.6$ mag saturate in the F775W and F850LP  filters ($I$ and $z$-band equivalent) of ACS/WFC. This roughly corresponds to masses at the hydrogen burning limit, assuming for the cluster a 1~Myr age and 414~pc distance \citep{menten2007}.  Concerning variability, the choreography of the HST Treasury Program observations did not enable a given field to be imaged in all filters in a single visit. Since most sources in the ONC present variability \citep{herbst2002}, which is typical for T Tauri stars for both accretion processes (Classical T Tauri Stars, CTTS) and rotation of spotted surface stars (Weak-line T Tauri stars, WTTS), this adds a  source of uncertainty, especially if the variability is wavelength dependent.

\begin{figure}
\plotone{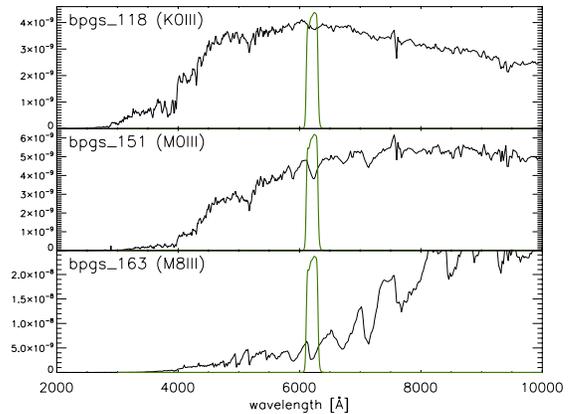}
  \caption{The medium-band MB\#620/17\_ESO866 filter response curve plotted over three standard spectra of giant stars (luminosity class III) of the Bruzual-Persson-Gunn-Striker catalog: \emph{46LMI} (spectral type K0), \emph{HD116870} (M0) and \emph{Z Cyg} (M8). The TiO band mapped by this filter is stronger for late spectral types.  \label{fig:bpgs_with_Rfilter}}
\end{figure}

To supplement the HST observations, ground-based data have been taken simultaneously from CTIO (near infrared) and La Silla (optical) on two nights in 2005 January. The Near Infrared J, H, and K data and results are presented in Robberto et al. (2009) \emph{in preparation}, and their discussion is not an aim of this work. At La Silla, we used the Wide Field Imager (WFI) at the ESO-MPI 2.2 m telescope to image the same field  in the $U$, $B$, $V$, and $I$ bands, as well as in the H$\alpha$ line and in a medium resolution filter at 6200 \AA. For stars earlier than  late K type, this filter samples the  R-band in a region free from major nebular lines, whereas for M-type stars and substellar objects it turns out be centered on a TiO absorption feature whose strength  depends on  the spectral type, growing towards the latest M types.  In Figure \ref{fig:bpgs_with_Rfilter} we show an example of three standard spectra taken from the Bruzual-Persson-Gunn-Striker spectral atlas \citep{gunn-striker83}, showing how the absorption band coincides with the filter profile.
A photometric study which includes narrow-band measurements at 6200 \AA\ might therefore be helpful for constraining the spectral type of cool stars, as well as disentangling the temperature-reddening degeneracy in the observed broad-band colors.
In what concerns the H$\alpha$ photometry, it can be used to assess membership,   and derive mass accretion rates through an absolute estimate of the accretion luminosity.

In this paper we present the WFI  data set. We describe the observations (\S\ref{section:observations}), their reduction and calibration (\S\ref{section:data_reduction}) and results \S\ref{section:results}.
In particular, the photometric catalog, the corresponding color-magnitude diagram and the completeness of our sample are described in  \S\ref{section:data}. In \S\ref{section:TiO} we focus on the $6200\AA$  medium band photometry, from which we derive a spectrophotometric index correlated with the spectral type. In \S\ref{section:halpha} we derive  the line excesses in the H$\alpha$ line. We summarize this work in \S\ref{section:conclusion}. This  data set allows to revisit  our current understanding of the ONC. This will be discussed in an accompanying paper (Da Rio et al. 2009, in preparation).

\section{Observations}
\label{section:observations}

The Wide Field Imager (WFI) is a focal
reducer-type camera mounted at the Cassegrain focus of the
2.2-m MPG/ESO telescope at La Silla. The camera features an array of 8 2k$\times$4k pixel CCDs, arranged in a 4$\times$2 mosaic. The optics deliver a field of view of 34 $\times$ 33 arcmin with 0.238 arcsec per pixel scale. The WFI CCDs have a typical read noise of 4.5 e$^-$ pix$^{-1}$ and a gain of 2.2~e$^-$ ADU$^{-1}$.

\begin{figure}
\plotone{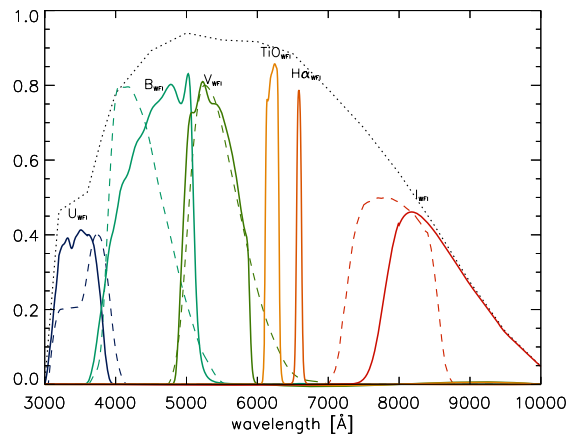}
 \caption{Band profiles for the six filters used in our observations (solid curves). The upper dotted line in black traces the CCD quantum efficiency; dashed lines in color represent the standard Johnson $U$, $B$, $V$ and Cousins $I_{\rm C}$ bands. The ESO879 $I$ band is centered at a significantly higher wavelength ($\sim8269$ \AA) than its Cousins counterpart. \label{fig:wfi_filters}}
\end{figure}

We selected the ESO877, ESO878, ESO843, and ESO879 broad-band filters, roughly corresponding to the standard $U$, $B$, $V$, and $I$ bands (note that the ESO879 $I$ band is centered at a wavelength higher than the standard $I_{\rm C}$ band, see Figure \ref{fig:wfi_filters}), and the medium-band MB\#620/17\_ESO866 and narrow-band  NB\#Halpha/7\_ESO856 filters.
We refer in this work to the  MB\#620/17 filter as the \emph{TiO band}.

\begin{deluxetable}{ccr}
\tablecaption{Central wavelength and FWHM of the WFI filters used.} 
\tablehead{\colhead{Band} & \colhead{$\lambda_{\rm cen}$} & \colhead{FWHM} \\
\colhead{} & \colhead{(\AA)} & \colhead{(\AA)} }
\startdata
$U$ & 3404 & 732 \\
$B$ & 4511 & 1335 \\
$V$ & 5395 & 893 \\
$I$ & 8269 & 2030 \\
TiO & 6208 & 194 \\
H$\alpha$ & 6588 & 74 \\
\enddata
\label{table:filters}
\end{deluxetable}

Our target field was observed in service observing mode on the nights of 2005 January 1 and 2 (hereafter indicated as Night A and Night B, respectively). The same observing strategy was adopted on both nights: for each filter we took 5 dithered exposures of 280~s (``Deep'') and 30~s (``Medium''), plus one single 3~s short exposure (``Short''). The five dithered exposures made it possible to eliminate the gaps between CCDs in the final mosaics, whereas the different exposure times increase the dynamic range of measured magnitudes. The final size of the mosaics, trimmed to the largest common area, is approximately 35 $\times$ 34 arcmin. Absolute photometric calibration was obtained by comparison with the Landolt SA 98 field \citep{landolt}, observed twice each night at low and high airmass.
The average seeing was better in Night A than in Night B (respectively 1.2\arcsec and 1.8\arcsec).

\section{Data reduction}
\label{section:data_reduction}
\subsection{Images and source catalogue}

The data were reduced using a version of {\sl alambic}, the ESO/MVM (Multi-Vision Model) image reduction system \citep{vandame04} as part of the EIS project for basic reduction of imaging data for a number of ESO imaging instruments.  To speed up the data processing, we installed the package and ran the pipeline parallel version on a server cluster at the Space Telescope Science Institute.

The package automatically performs all the common image reduction steps: removal of instrument effects (bias removal, flat field normalization), computation and removal of fringing  (only present in  \emph{I}-band exposures), bad-pixel-map creation, and normalization of gain levels between different CCD chips.

{\sl Alambic} also performs both relative and absolute astrometric calibration and distortion correction of the images to enable alignment and stacking of dithered frames. This step is achieved by retrieving an astrometric reference catalog, by default the GSC2 catalog, from the ESO archive. Since the GSC2 catalog turns out to be deficient in the Orion Nebula region, we used instead the 2MASS $J$-band catalog, which therefore represents our absolute astrometric standard. The near-IR\ 2MASS catalog does not trace well the distribution of sources at our  shortest wavelengths; thus, an iterative process has been carried out: the $J$-band reference catalog has been used for absolute astrometric calibration of TiO and $I$ bands. For the $V$ band we bootstrapped from these two bands, extracting from the TiO and $I$ calibrated images a secondary photometric catalog; the resulting $V$ catalog was then used to perform astrometric calibration of $B$ band images, and finally the $B$ catalog was used for the $U$ exposures. The final coordinates of bright blue sources detectable in all images have been checked to assess the accuracy of the procedure, which typically provides coordinates that agree to within one pixel between different bands. A better astrometric calibration is not needed for our purposes, considering that the average FWHM for the PSF in the data is $\sim 1.5$ arcsec, or  6 pixels.  After astrometric calibration, all final images with the same filter and exposure time have been coadded using SWarp, a package included in the \textsc{Terapix} package \citep{bertin02}.

For the four U,B,V,I broad bands point-spread function (PSF) photometry has been performed using the Daophot II package \citep{stetson87} on all the broad-band filter images. Unlike aperture photometry - that we used instead for the narrow band filters -, PSF photometry resolves and measures with good accuracy the luminosity of stars in highly crowded areas, such as the Trapezium region. We computed the PSF function for every image, using a set of reference stars located over the entire frame, and then refined this with an iterative sigma-clipping algorithm that rejected the candidates with excessive $\chi^2$.
PSF fitting was then performed on all the sources extracted with peak above $3\sigma$ of the local mean sky background, evaluated for every image using the SExtractor tool \citep{bertin02}. We did not attempt to recover  saturated sources, relying instead on the frames with shorter exposure times; only a small number of stars turned out to be saturated even in the 3s exposures too, and are therefore not present in our final catalog.

For each field and filter, we merged the photometric catalogs obtained with the different  exposure times on the basis of the absolute positions of the sources, using the \textsc{Daomaster} package \citep{stetson87}. Systematic differences in magnitude between different frames (mostly due to airmass variations) have been corrected assuming that the fundamental  instrumental magnitude is the one measured on the short (3s) exposures. The instrumental magnitudes of stars well measured in different images have been averaged.

For every filter and exposure time we eventually kept only the sources detected in both nights, in order to reduce contamination from spurious detections, which turns out to be not negligible due to the deep threshold (3$\sigma$)\ chosen for source identification.
We have matched the photometry in the six bands using again \textsc{Daomaster}, keeping only those that present a counterpart in two adjacent bands, starting from the reddest ones (i.e.,  sources found in $V$ and $I$; then $B$, $V$ and $I$; then $U$, $B$, $V$ and $I$). This because we expect the observed SEDs to increase with wavelength in the visible range for all the faint, red end of the population; in other words, faint sources may be present only at the longer wavelengths, remaining  invisible at the shortest ones; on the other hand, bright sources have been detected in all filters, if not saturated.
For the H$\alpha$ and TiO ($6200$\AA) bands we performed aperture photometry, evaluating and correcting the aperture correction to be applied using the brightest sources in the sample. These data have been first independently matched, keeping only stars detected in both filters on both nights to eliminate the eventual spurious detection, and afterwards have been measured and matched to the broadband catalog.

To ensure that our catalog is free from any additional spurious detections, we checked our final list of bona-fide candidates by visually inspecting their appearance  in every filter.
The data relative to the two nights have not been merged, thus in the final catalog the photometry of each source  has two entries. However, given the better sky conditions in Night A, we will concentrate our analysis to the data taken on this night.

\subsection{Absolute Calibration }
\subsubsection{Broad-band photometry}
Traditionally, the absolute calibration of optical photometry is achieved by determining the transformation between the instrumental and the Johnson-Cousins photometric systems. This method assumes a linear or polynomial dependance of the magnitude offset for a given band as a function of a color index. This relation should be derived from a number of standard stars observed under the same conditions.  In this work we chose not to perform a color correction, but instead we define a new photometric system based on the actual throughput of our passbands.   This turns out to be an advantage, considering that there are several sources of uncertainty regarding the color correction to be applied. The main reasons of concern are the following.

 First, as shown in Figure \ref{fig:plotCMD}, our set of observed PMS stars includes some with very high color indices, up to $V-I\sim6$. This is due to the combination of low \teff\ for the low mass objects and high $A_V$ in the region. The standard stars observed in the field SA$98$ reach only color terms up to $V-I\simeq2$. There is no reason to assume that the linear transformation extracted from these standard stars can be extrapolated to the very red PMS stars. In fact, typical standard stars do not provide any information about the color corrections at extremely red color.

\begin{figure}
\epsscale{1.1}
\plotone{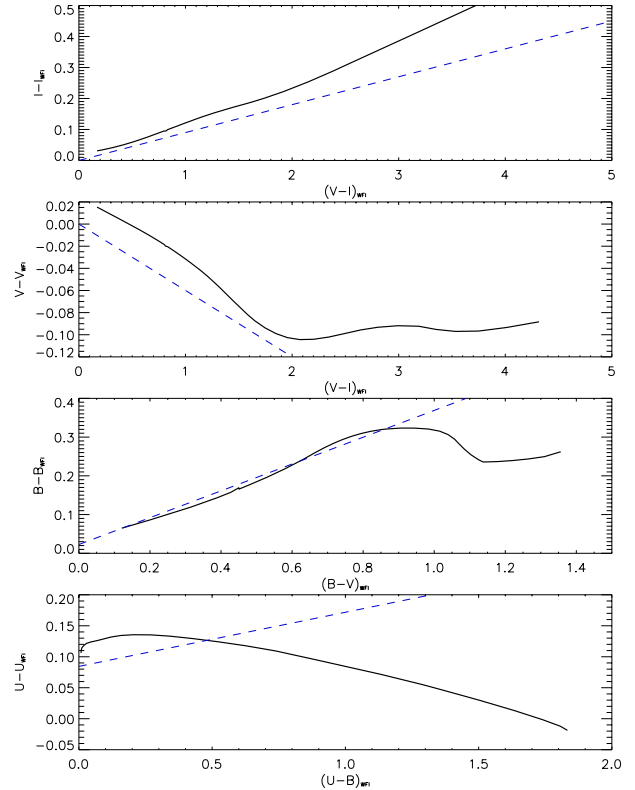}
\caption{Transformations between the WFI instrumental photometric system and the standard Johnson-Cousins system. The dashed line refers to the linear relations obtained from standard field observation (extrapolating the filters color terms comparing the observed colors with tabulated standard magnitudes spanning a more limited range in color); the solid line is the result of synthetic photometry using \textsc{NextGen} models with $\log g = 3.5$ and 2600 $\leq$ \teff\ $\leq$ 8200 K. $B$ and $U$ bands show high differences due to non linearity of the color correction. \label{fig:colorcorrections}}
\end{figure}
\begin{figure*}[ht!]
\epsscale{1.0}
\plotone{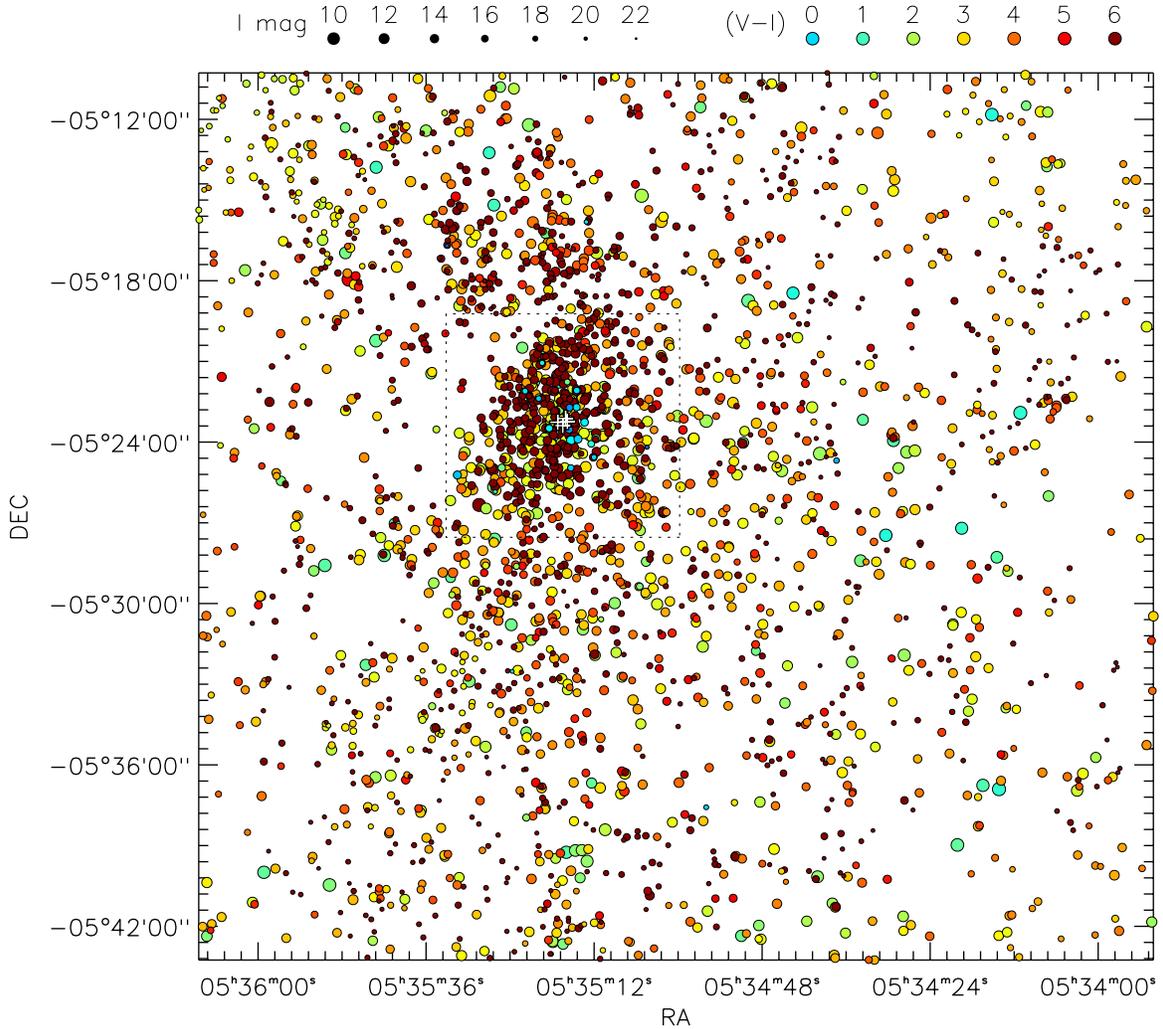}
\caption{Point-like sources selected in the final catalogue with at least $V$ and $I$ magnitude detected for both night A and B. The size of the dots increases with $I$ magnitude, and the colors represent the $V-I$ values as shown in the legend. The position of the four stars of the Trapezium cluster is shown with crosses. The dotted contour delimits the central area of 1pc of size centered on the star $\theta_{\rm 1}c$ of the trapezium cluster. The zoomed spatial distribution within such region is shown in Figure \ref{fig:plotradeccore}. \label{fig:plotradec}}
\end{figure*}

\begin{figure*}
\epsscale{1.0}
\plotone{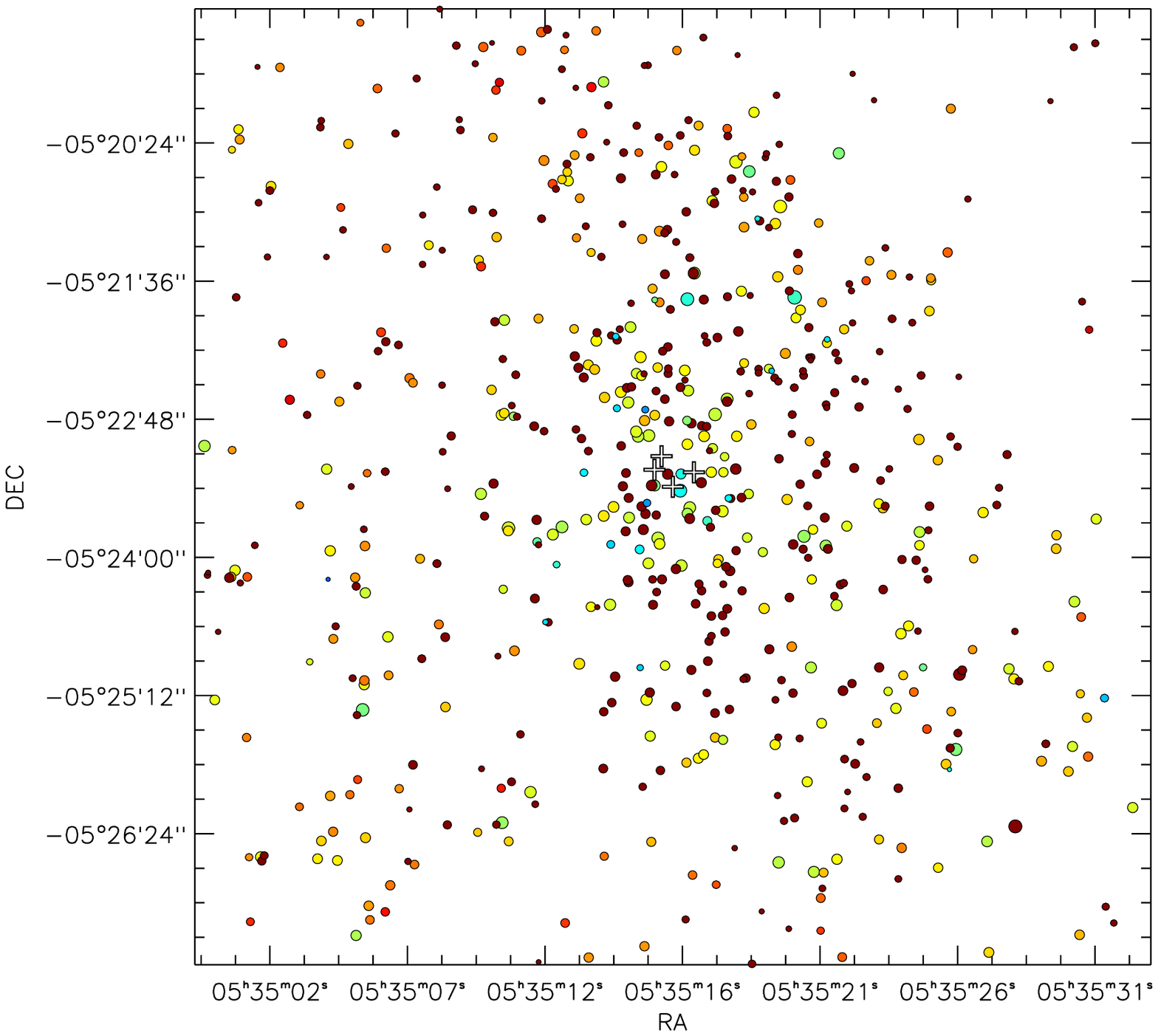}
\caption{Same as Figure \ref{fig:plotradec}, for the inner area (1~pc of size) centered on the star $\theta_{\rm 1}c$ \label{fig:plotradeccore}}
\end{figure*}

Second, in general the strengths of spectrum features that characterize stars of different \teff\ depends on several other parameters besides temperature, such as surface gravity and metallicity. In continuum-dominated spectral energy distributions, as for early-type stars, these parameters have little effect in the computation of the integrated colors, but for late-type stars, whose SEDs are increasingly dominated by broad, strong molecular absorption bands (mostly TiO and VO), magnitudes are sensitive to changes in the strength of these features resulting from differences in the physical parameters, and so are the colors and color corrections. Thus, besides our lack of knowledge about the shape of the color corrections for red objects (see above), these correction are not unique for all the ONC members.

In order to assess the first point we performed a test extracting synthetic photometry from a set of stellar atmosphere models. We used the \textsc{NextGen} models \citep{nextgenII}, choosing a set of spectra with 2600 $\leq$ \teff\ $\leq$ 8200 K and $\log g = 3.5$.\footnote{This value of $\log g$ is approximately one tenth of the solar value, and, according to the tables of H97 is a good enough approximation along all the temperature ranges in the ONC.}  Synthetic photometry has been derived using the \textsc{Gensynphot} software included in the \textsc{Chorizos} package
\citep{apellaniz}.
We defined the WFI bands by multiplying the filter transmissions by the CCD quantum efficiency curves, quantities that are well known for the WFI instrument, and then computed the zero-points of this newly defined \textsc{VegaMag} photometric system using the calibrated Vega spectrum distributed with the \textsc{Chorizos} package.
The measured instrumental magnitudes, therefore, are already expressed in this photometric system except for constant zero-points depending on the observing conditions. These offsets in fact coincide with the zero-points derived from standard field observations according to the standard method for photometric calibration (when we consider the color term relative to the calibrated magnitudes of the standard stars). This fact is clear considering the standard transformation, for instance, to calibrate an instrumental B-band photometry. We have that $(B-B_{WFI,{\rm Ins}})=ZP+CC\cdot(B-V)$, where $B_{WFI,{\rm Ins}}$ is corrected in order to be valid for the airmass of the science observation in this band, from the two repetitions of standard field observations which were originally taken at low and high airmass each night. The intercept $(B-V)=0$ corresponds to a Vega-type star, and in this case we have that $B-B_{WFI,{\rm Ins}}=ZP$. Furthermore, in the VegaMag convention, all the colors must be zero for a Vega-type star, so $(B-B_{WFI,{\rm VegaMag}})=0$. From these two expressions, finally, $(B_{WFI,{\rm VegaMag}}-B_{WFI,{\rm Ins}})=ZP$. The same method used for all the bands results in an absolute calibration of the photometry in the instrumental-VegaMag photometric system.

We then computed the synthetic magnitudes for stars with different \teff\ (i.e., different colors) in both the instrumental and the Johnson-Cousins photometric systems, determining the color-correction relations. The results are shown in Figure \ref{fig:colorcorrections} for the four broad-band filters used. In every case the non-linearity of these transformations is evident for high color terms; this effect affects dramatically the $B$ band, with shifts up to few tenths of magnitude from the linear color correction relation at high $B-V$.

While in principle it could be possible to apply the simulated color correction law, these are sensitive not only to the stellar parameters, but also to the accuracy of the synthetic atmosphere models. Unfortunately the uncertainties present in all these grids, especially in the temperature range of the M-type stars, are still limiting factors for an accurate evaluation of the integrated colors \citep[see][]{baraffe08}. Therefore, it is clear that keeping our catalog in the instrumental system removes most of the uncertainties related to the absolute photometric calibration, and, as we are going to show in a second paper within this project, preserves the capability of studying the stellar parameters through direct comparison of the observations with theoretical models. Our treatment of the color effects at high color terms and low surface gravity (in both band transformations and bolometric correction) represents a major improvement over the H97 analysis.

\subsubsection{Narrow-band photometry}
\label{section:narrowbandcalibration}
Due to the lack of standard star reference photometry in the two narrow bands in our dataset (the 6200\AA\ TiO band and the H$\alpha$ band), a traditional zero-point calibration approach cannot be followed; thus, we calibrated these data as follows.

First, the photometry of the two nights has been cross-calibrated, matching identical stars and determining the global average offset of night B versus night A, this step allows us to clean the sample from spurious detections. Then the magnitudes have been converted to units of flux (counts s$^{-1}$). We computed the ratio between the actual efficiencies (measured ADU / flux) between the TiO band and the H$\alpha$ filter, taking into account the differences in the filter passbands, the quantum efficiency $QE(\nu)$, of the WFI CCDs, airmass  -- deriving the  atmospheric extinction at 6200 \AA\ linearly from the ones measured in $V$ and $I$ -- and photon energy. In this way we derived \emph{a priori} that, considering the observations of night A, a constant flux observed in both filters (in terms of Jy) corresponds to a number of ADUs in the H$\alpha$ images $\sim2.70$ lower that in the TiO images. Therefore, correcting for this factor, we {\em relatively} calibrate the H$\alpha$ photometry on the TiO one. The subsequent absolute calibration of the latter, therefore, calibrates the H$\alpha$ as well.

The problem is therefore to find the magnitude offset to apply to the TiO photometry in order to express our data in standard VegaMag. By definition this offset is such that a for a Vega-type star all the colors are zero, for instance $({\rm V_{WFI}-TiO})=0$ or $({\rm TiO-I_{WFI}})=0$. Unfortunately our sample lacks a significant number of such objects, requiring to use stars of later type to improve the precision of the calibration. If we consider the most abundant M-type stars in our catalog, however, the behavior of the measured flux in TiO is strongly affected by the absorption feature shown in Figure \ref{fig:bpgs_with_Rfilter}. We consider therefore the intermediate temperature stars (G and early K spectral types), isolated in our sample by matching our photometric catalog with the spectroscopy of \citet{hillenbrand97}. A first order estimate of the calibration to be applied to the TiO photometry could be to impose that, for these stars, the magnitude at 6200nm is equal to the linear interpolation of the V and I magnitudes, assumed at a wavelength equal to the average wavelength of the band profiles (see Table \ref{table:filters}). We investigate the correctness of this assumption using synthetic photometry on a grid of atmosphere models. We consider the NextGen \citep{nextgenII} grid, for solar metallicity and surface gravity $\log(g/g_\odot)=3.5$, an average value for PMS stars. In the temperature range we selected - for which the uncertainty of the theoretical spectra is negligible \citep{baraffe08} -, the models predict that the TiO magnitude is on constantly 0.17 mag brighter than the assumed linear interpolation, and therefore we applied this offset in order to calibrate the TiO photometry. Once the photometry in VegaMag is calibrated, we are able to express it into physical flux, calculating the zero-point of the band of $3.175150\cdot10^6$ mJy by means of integration of a reference Vega spectrum using the \emph{STSDAS Synphot} software. Given that we calibrate the H$\alpha$ photometry in units of flux relatively to the TiO, the same zero point, as explained above, is therefore the same also for the absolute calibration of the H$\alpha$ photometry.\\

\section{Results}
\label{section:results}

\subsection{The Source Catalog}
\label{section:data}
The final catalog includes 2,621 point-like sources detected in $I$ band, 1,523 of which have both $V$ and $I$ magnitudes, 1,134 also $B$ magnitudes and 431 have $U$ magnitudes. Narrow band photometry is available for 1040 stars. As mentioned above, the catalog itself provides for each star 2 distinct photometric values for nights A and B, respectively; however, given the poor sky conditions of night B, we present only the results from the observations during the first night.

In Figure \ref{fig:plotradec} we present the distribution on the sky of the detected sources. The general structure of the cluster, elongated in the north-south direction as already described in \citet{hillenbrandhartmann98}, is evident. In this figure we also highlight different luminosities with the dot size and the $ (V-I)$ indexes with colors.

\begin{figure*}
\plotone{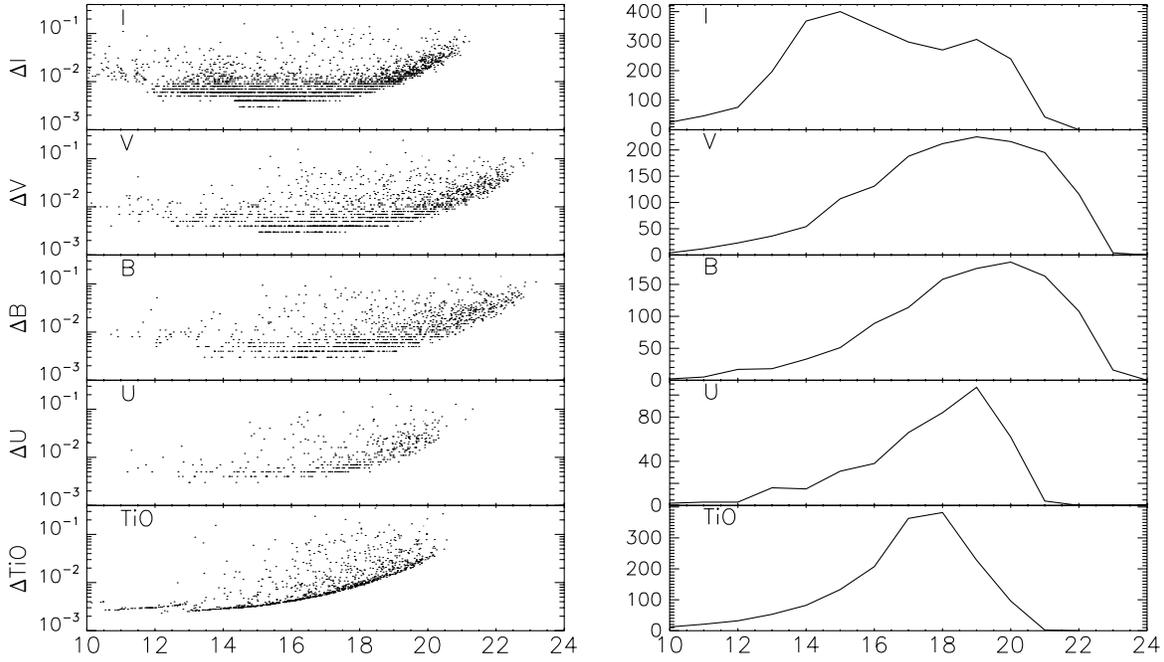}
\caption{{\em Left panel}: photometric errors for the five bands observed. {\em Right panel}: luminosity functions. The double peak in the I band is reasonable due to contamination by background stars. \label{fig:photometricerrors}}
\end{figure*}

Figure \ref{fig:photometricerrors}a shows the photometric errors as a function of luminosities for the UBVI+TiO filters.
The visible multiple trends in the distribution of photometric errors
are due to the superposition of deep, medium and short exposures. Bright stars are likely to be measured only in short exposures being saturated in the long ones,  and their photometric error turns out relatively high. Figure \ref{fig:photometricerrors}b presents the observed luminosity functions in the 5 bands.

\begin{figure}
\plotone{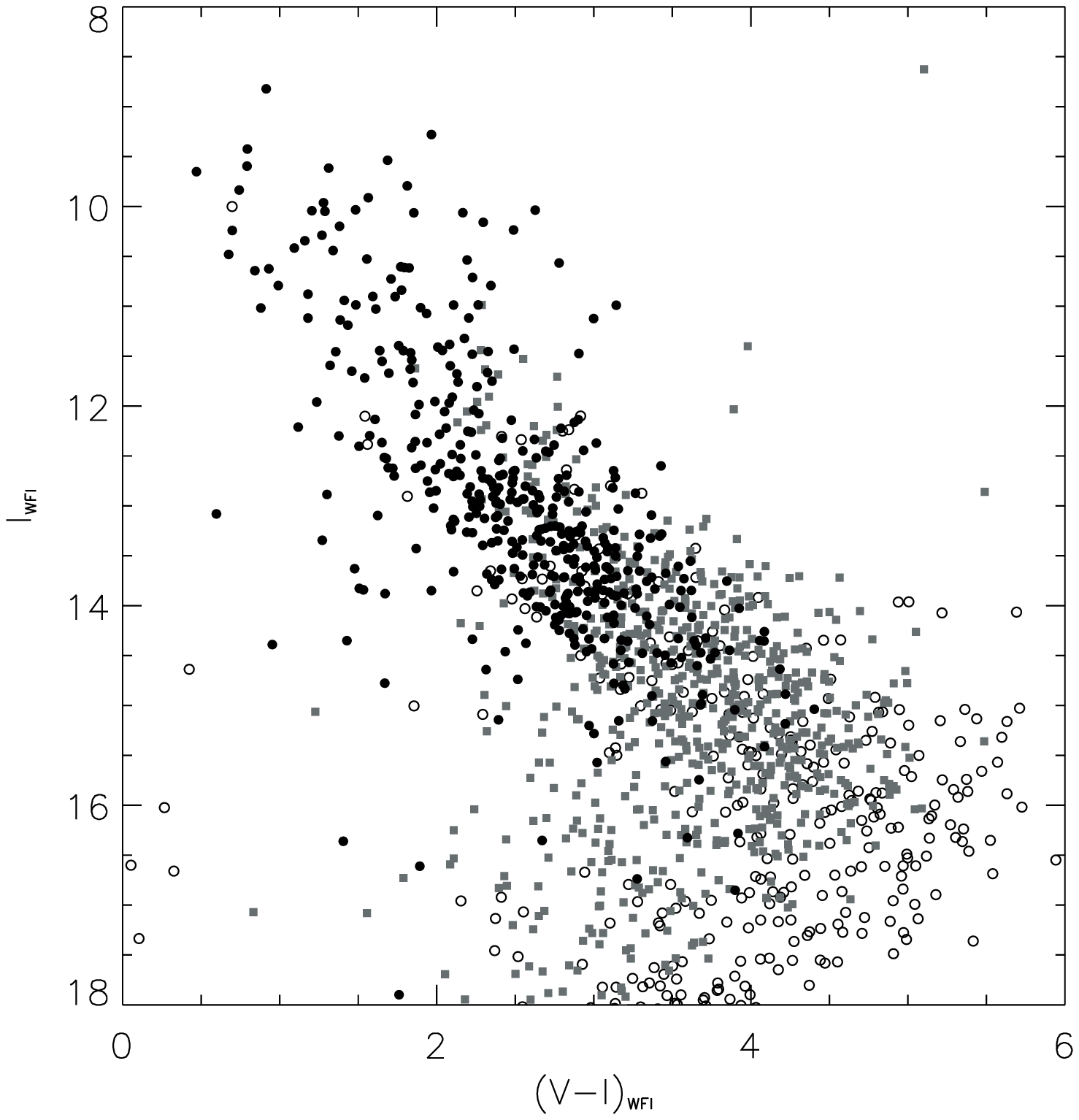}
\caption{Instrumental color-magnitude diagram for the sources present in our catalog: filled circles are objects with all $U$, $B$, $V$, and $I$ measures, squares are sources lacking $U$, and open circles are objects with only $V$ and $I$ magnitudes. \label{fig:plotCMD}}
\end{figure}

In Figure \ref{fig:plotCMD} we present the  color-magnitude diagram for the four broad-band filters. We use different symbols for the stars having all the four broad-band magnitudes are available, only $B$, $V$, and $I$, or only $V$ and $I$. Given the red colors of the ONC members, the faint stars lack signal at the shorter wavelengths.

\begin{figure}
\plotone{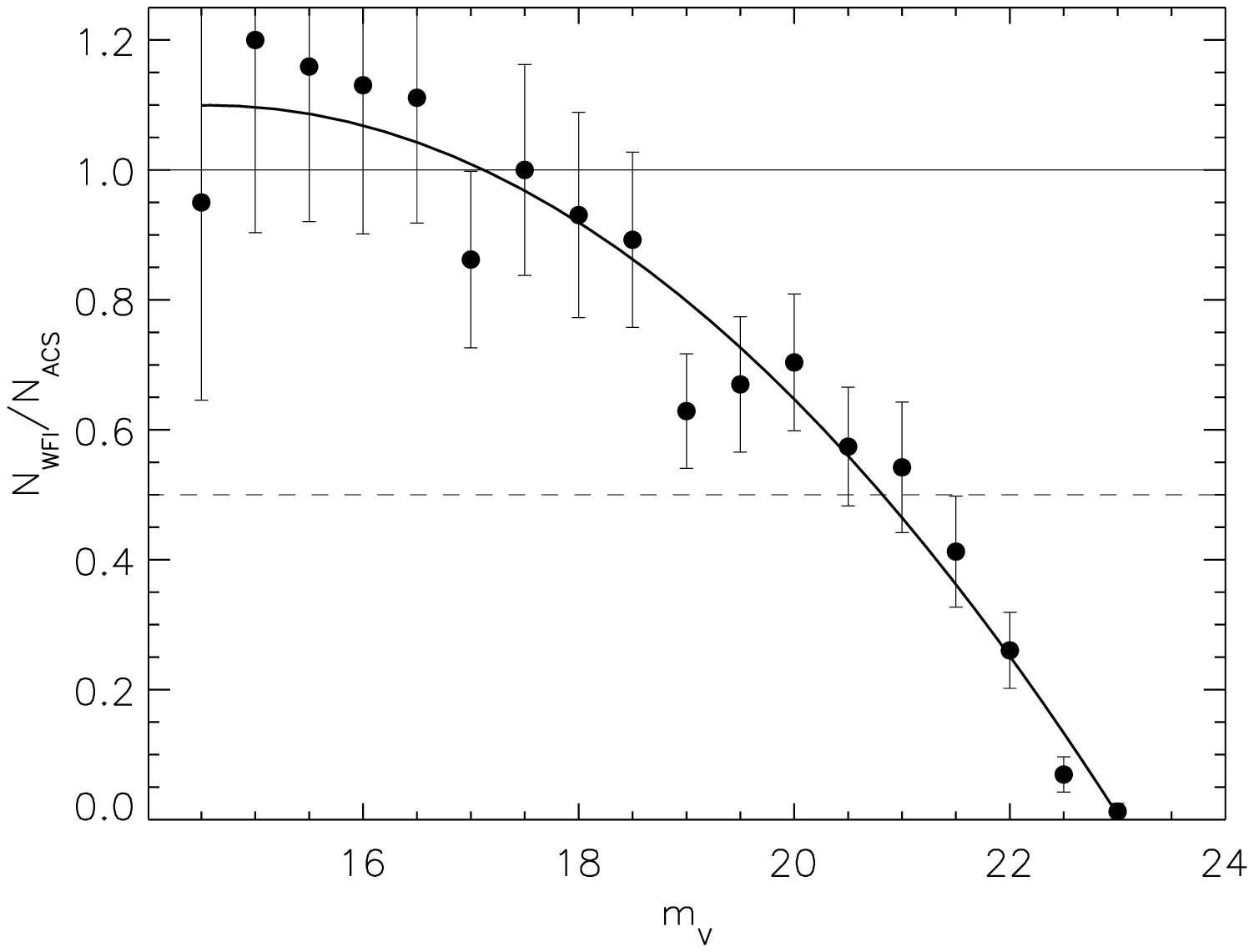}
\caption{Completeness function for the V band. Dots represent the measure ratio between the number of sources counted in bins spaced 0.5 magnitudes in the WFI and the HST/ACS catalog respectively, considering only the common FOV of the two surveys. A polynomial fit of the second order is overlayed. The apparent systematic completeness greater than unity in the bright end is due to source saturation of the ACS imaging.  \label{fig:completeness_with_ACS}}
\end{figure}

We estimated the completeness limit of our survey, comparing our luminosity functions (LFs) in V band with the one derived from the HST/ACS catalog (Robberto et al. 2009, in preparation) in the V- equivalent filter F555W.
The HST survey is characterized by a considerably fainter detection limit. In order to allow for a consistent comparison, we considered only sources present in the common field of view. This, which is about $55\%$ of the $\sim1100$ square arcmin total WFI FOV and contains about $70\%$ of the WFI sources we detected, is wide enough to be well representative of the whole observed area, covering both the central part of the cluster, more affected by crowding, and the loose periphery.  The ratio between LF counts, as a function of the V magnitude, directly provides our completeness function  in this band. It is shown in Figure \ref{fig:completeness_with_ACS}. At the bright end of the distribution the deep ACS catalog is deficient of sources with respect to WFI,  due to saturation and to the presence of non stellar sources (e.g proplyds or circumstellar halos) which may contaminate our lower resolution of ground based catalog. At the faint end, the ratio of sources detected with WFI to $\sim 50\%$ at $V\simeq 20.8$, which we therefore consider as representative of our completeness limit for the source detected in both V- and I-band.

\begin{deluxetable*}{rrrrrrrrrrrrrr}
\tablewidth{0pt} \tablecaption{WFI photometric catalog relative to Night A observations.
\label{table:catalog}}
\tablehead{\colhead{ID} & \colhead{RA} & \colhead{Dec} & \colhead{U} & \colhead{$\Delta$U} & \colhead{B} & \colhead{$\Delta$B} & \colhead{V} & \colhead{$\Delta$V} & \colhead{TiO} & \colhead{$\Delta$TiO} & \colhead{I} & \colhead{$\Delta$I} & \colhead{H97 ID} \\
\colhead{} & \colhead{(J2000.0)} & \colhead{(J2000.0)} & \colhead{} & \colhead{} & \colhead{} & \colhead{} & \colhead{} & \colhead{} & \colhead{} & \colhead{} & \colhead{} & \colhead{} & \colhead{}}
\startdata
   1 &    05~35~47.01 &    -05~17~56.9 &  \nodata &  \nodata &   16.093 &    0.004 &   13.728 &    0.004 &   12.097 &    0.003 &    8.626 &    0.027 &            992 \\
   2 &    05~35~20.71 &    -05~21~44.4 &    9.886 &    0.008 &   10.041 &    0.009 &    9.736 &    0.008 &    9.540 &    0.004 &    8.822 &    0.021 &            660 \\
   3 &    05~35~05.20 &    -05~14~50.3 &   13.260 &    0.007 &   12.141 &    0.011 &   11.246 &    0.017 &   10.535 &    0.003 &    9.280 &    0.036 &            260 \\
   4 &    05~34~14.16 &    -05~36~54.1 &   11.081 &    0.005 &   10.697 &    0.009 &   10.219 &    0.010 &    9.929 &    0.004 &    9.425 &    0.016 &        \nodata \\
   5 &    05~35~21.31 &    -05~12~12.7 &   13.057 &    0.004 &   12.043 &    0.052 &   11.225 &    0.015 &   10.646 &    0.003 &    9.538 &    0.018 &            670 \\
   6 &    05~35~28.41 &    -05~26~20.1 &  \nodata &  \nodata &  \nodata &  \nodata &  \nodata &  \nodata &    9.601 &    0.004 &    9.552 &    0.016 &            831 \\
   7 &    05~34~11.11 &    -05~22~54.6 &   11.199 &    0.014 &   10.900 &    0.008 &   10.388 &    0.010 &   10.130 &    0.004 &    9.596 &    0.018 &        \nodata \\
   8 &    05~34~49.97 &    -05~18~44.6 &   12.583 &    0.004 &   11.496 &    0.010 &   10.928 &    0.010 &   10.474 &    0.004 &    9.616 &    0.023 &            108 \\
   9 &    05~35~16.72 &    -05~23~25.2 &    9.928 &    0.021 &   10.075 &    0.026 &   10.122 &    0.012 &    9.908 &    0.006 &    9.652 &    0.019 &        \nodata \\
  10 &    05~34~39.75 &    -05~24~25.6 &   14.065 &    0.004 &   12.535 &    0.010 &   11.606 &    0.008 &   10.906 &    0.003 &    9.794 &    0.021 &             45 \\
\nodata & \nodata & \nodata & \nodata & \nodata & \nodata & \nodata & \nodata & \nodata & \nodata & \nodata & \nodata & \nodata & \nodata \\
\enddata
\tablecomments{Full table available as on-line data. The H$\alpha$ photometry is presented separately in Table \ref{table:haexcess}.  All the magnitudes are given the \textsc{VegaMag} instrumental system. The last column reports the corresponding ID of the source in the catalog of \citet{hillenbrand97}, if present.}
\end{deluxetable*}

Our catalog, presenting for each star coordinates, calibrated magnitudes and photometric errors in $U$, $B$, $V$, $I$, and TiO band is given in Table \ref{table:catalog}  (including only photometry derived from night A observations, used in this work). We matched our catalog with the one of \citet{hillenbrand97}, and the last column of  Table \ref{table:catalog} reports (if present) the ID number of the latter.

In principle, the photometry from the second night could have been used to estimate the stellar variability and to provide a better characterization of the uncertainty associated to photometry. However, the poor seeing conditions of the second night strongly affect also the accuracy of the photometry. This because in a field like the Orion Nebula source confusion and non uniformity of the nebular background have a stronger influence on the photometric accuracy when the PSF is larger. If we limit ourselves to the brightest part of the photometric catalog, where these uncertainties have lower impact, we see that the standard deviation of the difference in magnitudes between the two night, for the bright $50\%$ of the population, is 0.17 mag in the I-band, 0.30 mag in V-band and B-band, 0.35 mag in U-band, on a time baseline of about 24 hours. This is compatible with what found by \citet{herbst2002},  who report that more than $50\%$ of the cluster members display a I-band variability larger than 0.2 magnitudes over periods of a few days.

The reported coordinates are the positions measured on the reduced images in I band, whose field distortion has been corrected by means of a polynomial warping using the 2mass astrometry as a reference for all the matched sources. We estimate the precision of our astrometry comparing our positions with those of the H97 catalog and with the HST/ACS photometry. In the comparison with H97, the standard deviation of the coordinate offsets is $0.24\arcsec$ in RA and $0.22\arcsec$ in Dec, while with respect to the HST photometry they decrease to $0.13\arcsec$ and $0.11\arcsec$ respectively. Given that the astrometric precision of the ACS catalog is much less than 1 pixel ($0.05\arcsec$), the latter dispersions are a good estimate of the uncertainty of our astrometry. Therefore, we consider the astrometric precision of our WFI catalog to be $\sim0.13\arcsec$, which therefore improves by a factor of 2 over H97.

The H$\alpha$ catalog, expressed in terms of non photospheric flux excess, is presented independently (see \S\ref{section:halpha}).

It is worthwhile to estimate the improvement in the optical characterization of ONC stars of our sample with respect to the one of H97. The two catalogs cover a similar extension on the sky, but the latter, counting 1576 sources with respect to our 2621 stars, is centered on the Trapezium cluster whereas our observations are centered about 6$\arcmin$ to the south-west. If we consider only stars detected in the common area of the two catalogs, we have 2246 stars from WFI data and 1463 from H97, 926 of which with known spectral types. We highlight that in the presented catalog we don't recover all the H97 sources in the common area, but we lose about 90 stars. This is mostly due for saturation in the WFI observations, but we also found for small sparse groups of stars evidences of  systematic shifts in the coordinates between our coordinates and the H97 ones, large enough to fail our matching algorithms and probably due to erroneous coordinates in the literature. We don't consider those stars among the matched ones (last column of Table \ref{table:catalog})

\subsection{The TiO spectrophotometric index}
\label{section:TiO}
The depth of the TiO spectral feature shown in Figure \ref{fig:bpgs_with_Rfilter} can be used to classify late-type stars. To this purpose, we have defined a index, [TiO], using the $V$, $I$ and TiO luminosities.

Our [TiO] index represents, roughly speaking, the ``lack'' of flux, in  magnitudes, measured in the TiO feature, and is defined as the difference between the linear interpolation at 6200 \AA\ of the $V$ and $I$ magnitudes and the apparent magnitude in the TiO band:
\begin{eqnarray}
 {\TiO}& =& m_{\rm TiO,interp} - m_{\rm TiO} \nonumber \\
&    = & \bigg(V_{\rm WFI}-(V-I)_{\rm WFI}\cdot\frac{\lambda_{\rm TiO}-\lambda_V}{\lambda_I-\lambda_V}\bigg) - m_{\rm TiO}   \end{eqnarray}
where $\lambda_V$, $\lambda_I$ and $\lambda_{\rm TiO}$ are the central wavelengths of the three filters, as reported in Table \ref{table:filters}.

\begin{figure}
\plotone{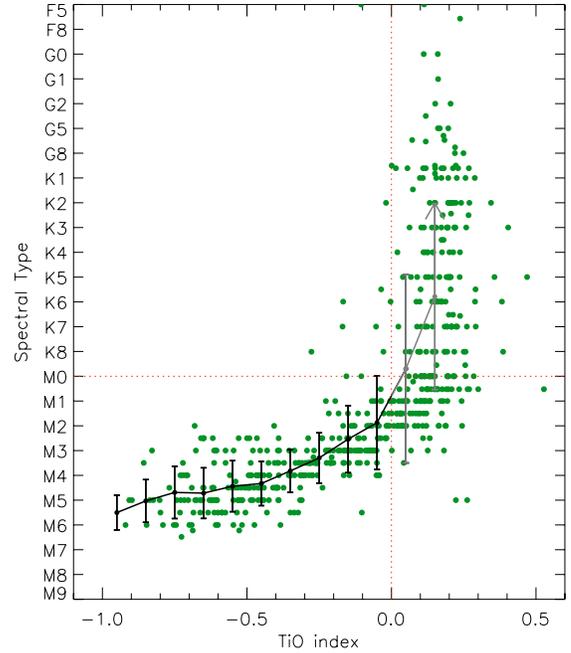}
\caption{The spectral type vs. [TiO] index relation showing all the 583 stars for which H97 spectral types and TiO photometric index are available, 298 of which with [TiO]$<$0. The law we derived for M-type stars with low photometric errors ($\Delta m_{\rm TiO}\leq 0.03$) is overplotted (black line).
   \label{fig:tio_index_laws}}
\end{figure}

In Figure \ref{fig:tio_index_laws} we plot our [TiO] index versus the spectral type extracted for all the stars available from \citet{hillenbrand97}. This sample counts 583 stars. For M-type stars, the depth of the TiO feature measured by means of narrow-band photometry increases moving towards higher types, and is up to one magnitude fainter than the linear interpolation between $V$ and $I$. In this spectral range, therefore, there is a definite correlation between [TiO] and spectral type, whereas for earlier classes it assumes a roughly constant value of $\simeq 0.1 - 0.2$ mag. This limits its use to M type stars and possibly brown dwarfs, which are not included in the H97 list.

By isolating within our catalog the stars with small photometric errors ($\Delta m_{\rm TiO}<0.03$), we derived a quantitative relation between the TiO index and spectral type, extracted a moving average, as shown in Figure \ref{fig:tio_index_laws}, and shown in Table \ref{table:tiorelation}. The 1$\sigma$ uncertainty is also reported, showing that our relation is accurate to better than 1 spectral sub-class for M3-M6 types and better than 2 spectral sub-classes for M0-M2 types.

\begin{figure}
\plotone{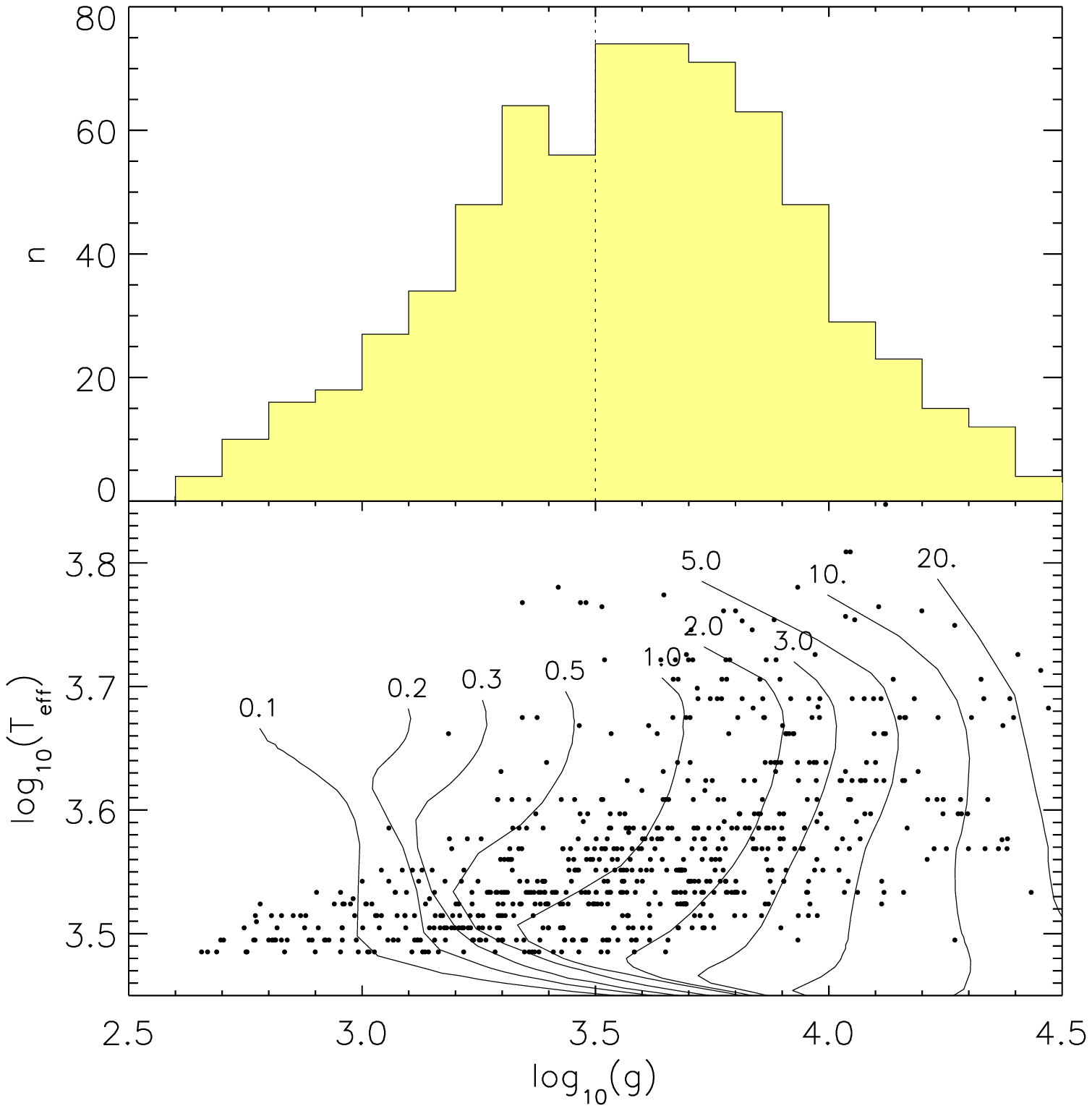}
\caption{{\em Upper panel:} the distribution of $\log g$ (cgs units) from \citet{hillenbrand97}, limited to the sample of stars used to derive our [TiO] - spectral type relation. {\em Lower panel}: $T_{\rm eff}$ vs $\log g$ from the same work. Isochrones from 0.1~Myr to 20~Myr from evolutionary models of \citet{dantona-mazzitelli94} are overlayed highlighting that the derived broad distribution of $\log g$ comes from the interpretation of the H-R diagram spread as due to an age distribution.
   \label{fig:logg_distrib}}
\end{figure}

We investigated the sources of scatter evident in the figure looking for dependencies on other parameters. While assuming equal metallicity for the ONC members is quite legitimate, an age spread for the ONC members leads to a distribution of surface gravity for stars of a given mass, given that the stellar radius changes during PMS evolution.
Specifically, we consider the tabulated values of $\log g$ derived in \citet{hillenbrand97} using \citet{dantona-mazzitelli94} evolutionary models on their H-R diagram. The latter was obtained from spectral types converted into $T_{\rm err}$ according to \citet{cohencuhi79} and bolometric corrections applied to reddening corrected V-I photometry. In Figure \ref{fig:logg_distrib} we present the surface gravity distribution, and presenting the H97 stars in the $T_{\rm eff}$ vs $\log g$ in comparison with PMS isochrones of different ages, we show that the measured age spread in the ONC is the main cause of a wide distribution of surface gravity in our sample.

\begin{figure}
\plotone{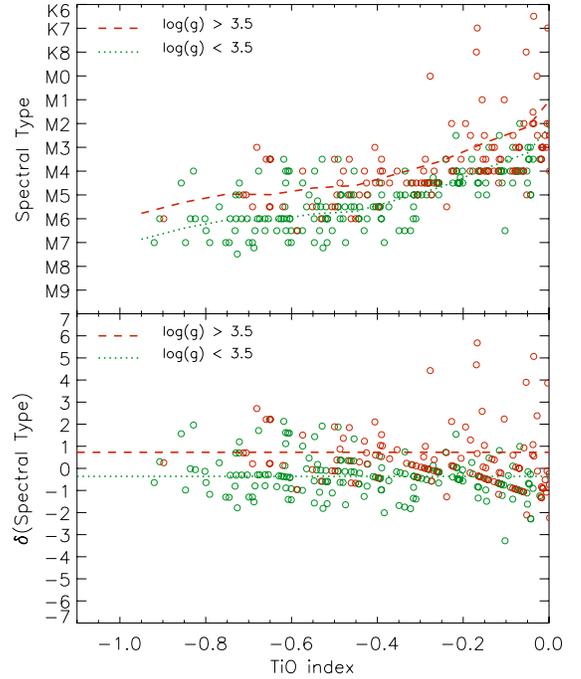}
\caption{\emph{Upper panel:} the relation for values of TiO indices for the case of $\log g > 3.5$ and $<3.5$ in cgs units. \emph{Bottom panel:} the displacement in units of spectral types with respect to the global dependence of Table \ref{table:tiorelation}, showing the average in the 2 cases. \label{fig:TiOindex_logg_test}}
\end{figure}

We divided our stars in two samples, of low and high surface gravity, and we found that variations in surface gravity introduce a systematic offset in the TiO index vs. spectral type relation.
 In Figure \ref{fig:TiOindex_logg_test} we show the result of isolating stars with $\log g <3.5$ and $\log g >3.5$ in cgs units (the threshold being the average value of $\log g$ in our sample). We subtracted the "predicted" spectral type from the [TiO] index using our law tabulated in Table \ref{table:tiorelation} to the "actual" one, and for each of the two subsamples we computed the average of these residuals (Figure \ref{fig:TiOindex_logg_test}, bottom panel). This is not zero in the two cases, suggesting a real dependance of the behavior of our index on the stellar surface gravity.

We then carried out a statistical test of the average of two subsets, assuming a gaussian distribution, and comparing the shift between the two mean values from zero with the relative rms. In other words, we computed the probability $P$ that the two distributions are statistical representations of the same distribution considering the estimator $z$ as follows:
\beq
 z=\frac{m_1-m_2}{\sqrt{\frac{\sigma_1^2}{N_1}+\frac{\sigma_2^2}{N_2}}}
\eeq
where $m_1-m_2=1.1$ spectral sub-types is the difference between the two mean values (shown in Figure \ref{fig:TiOindex_logg_test}b by the dashed and dotted lines), $\sigma_1$ and $\sigma_2$ are the standard deviations of the two sets, and $N_1$ and $N_2$ are the total number of stars in each set. We derived $z=5.9$, corresponding to a probability (for the gaussian hypothesis) of about $2\cdot10^{-9}$. This proves that the behavior of the TiO index is dependent on surface gravity, and, for a given value of the index, the higher the surface gravity, the lower the derived spectral type. The difference of the averages of $\log g$ in the two subsets is equal to $0.53$ dex. We derive a shift in the [TiO] vs $\log T_{\rm eff}$ relation of 2 subtypes per unit of $\log g$ with respect to our empirical law tabulated in Table \ref{table:tiorelation}, valid for the average value $\log g=3.5$.

\begin{deluxetable}{rrr}
\tablewidth{0pt}
\tablecaption{Empirical TiO vs. spectral type relation}
\tablehead{
\colhead{TiO index} & \colhead{Sp Type} & \colhead{$\sigma$ [types]}\\}
\startdata
-0.95 & M5.5 & 0.7 \\
-0.85 & M5.0 & 0.9 \\
-0.75 & M4.7 & 1.1 \\
-0.65 & M4.7 & 1.0 \\
-0.55 & M4.4 & 1.0 \\
-0.45 & M4.3 & 0.9 \\
-0.35 & M3.8 & 0.9 \\
-0.25 & M3.3 & 1.0 \\
-0.15 & M2.5 & 1.4 \\
-0.05 & M1.9 & 1.9 \\
0.05 & $<$M3.5 & \nodata \\
0.15 & $<$M0.5 & \nodata \\
\enddata
\label{table:tiorelation}
\end{deluxetable}

The dependence of the [TiO] index on surface gravity is intriguing, but its interpretation is not necessarily trivial. In fact, as the depth of the TiO feature is influenced by stellar surface gravity, also the behavior or the V- and I- band of cold stars depend on the luminosity class of the star, and therefore on the stellar parameter $\log g$. Moreover, the younger is a low-mass PMS star, the lower is its surface gravity (because of its larger radius); the dependance we found could be, therefore, introduced by different ages in the considered mass range. For instance, mass accretion, which is typically higher for young pre-main sequence objects introduces veiling in the stellar spectrum, which decreases the relative depth of the spectral absorption features, causing a star of a given temperature to have a  shallower TiO feature at younger ages.

In any case, neither age nor surface gravity is known for stars of the ONC for which the spectral type is not determined. This implies that using our [TiO] - spectral type relation to classify M-type stars cannot be refined by the aforementioned consideration about surface gravity. Therefore, we leave a detailed analysis of the sources of this effect in the [TiO] index to a future paper, considering, for the moment, only the global relation expressed in Table \ref{table:tiorelation}.

Using this, we compute the value of [TiO] for all  stars not included in the H97 spectral catalog for which our V,I and TiO band photometry is available. Selecting only sources with [TiO]$<0$. We derive the spectral type and its uncertainty for 217 additional stars. The results are tabulated in Table \ref{table:tiostars}.

\begin{longtable}{rrrrrrr}
\tablecaption{New M-type stars classified from the [TiO] index}
\tablehead{\colhead{} & \colhead{ID} & \colhead{RA} & \colhead{Dec} & \colhead{[TiO]} & \colhead{Spectral} & \colhead{$\sigma$} \\
\colhead{} & \colhead{*} & \colhead{(J2000.0)} & \colhead{(J2000.0)} & \colhead{(mags)} & \colhead{Type} & \colhead{} }
\startdata
 1 & 1306 & 05~36~00.49 & -05~41~10.9 & -0.145 & M2.5 & 1.4 \\
 2 & 608 & 05~35~59.48 & -05~37~09.6 & -0.478 & M4.4 & 0.9 \\
 3 & 1642 & 05~35~59.25 & -05~33~00.4 & -0.541 & M4.4 & 1.0 \\
 4 & 630 & 05~35~57.51 & -05~39~51.3 & -0.080 & M2.1 & 1.7 \\
 5 & 1571 & 05~35~57.35 & -05~25~15.6 & -0.457 & M4.3 & 0.9 \\
 6 & 758 & 05~35~55.97 & -05~42~26.3 & -0.225 & M3.1 & 1.1 \\
 7 & 1084 & 05~35~54.49 & -05~26~45.6 & -0.241 & M3.2 & 1.0 \\
 8 & 1370 & 05~35~52.90 & -05~25~43.9 & -0.068 & M2.0 & 1.8 \\
 9 & 732 & 05~35~52.20 & -05~39~24.7 & -0.001 & M0.8 & 2.8 \\
 10 & 746 & 05~35~51.76 & -05~17~39.4 & -0.393 & M4.0 & 0.9 \\
 11 & 815 & 05~35~48.71 & -05~42~14.4 & -0.369 & M3.9 & 0.9 \\
 12 & 660 & 05~35~47.65 & -05~37~38.8 & -0.322 & M3.7 & 0.9 \\
 13 & 304 & 05~35~46.42 & -05~41~00.7 & -0.400 & M4.1 & 0.9 \\
 14 & 1233 & 05~35~44.58 & -05~32~55.9 & -0.321 & M3.7 & 0.9 \\
 15 & 515 & 05~35~43.44 & -05~40~55.0 & -0.049 & M1.9 & 1.9 \\
 16 & 938 & 05~35~42.66 & -05~40~42.2 & -0.462 & M4.3 & 0.9 \\
 17 & 1404 & 05~35~42.61 & -05~26~33.7 & -0.550 & M4.4 & 1.0 \\
 18 & 778 & 05~35~41.30 & -05~38~32.9 & -0.382 & M4.0 & 0.9 \\
 19 & 890 & 05~35~41.23 & -05~38~29.4 & -0.410 & M4.1 & 0.9 \\
 20 & 999 & 05~35~41.13 & -05~42~51.1 & -0.472 & M4.4 & 0.9 \\
 21 & 379 & 05~35~40.83 & -05~32~01.8 & -0.311 & M3.6 & 0.9 \\
 22 & 1258 & 05~35~37.06 & -05~37~37.2 & -0.569 & M4.5 & 1.0 \\
 23 & 1390 & 05~35~36.55 & -05~40~25.1 & -0.510 & M4.4 & 1.0 \\
 24 & 562 & 05~35~36.35 & -05~31~37.8 & -0.461 & M4.3 & 0.9 \\
 25 & 629 & 05~35~35.95 & -05~38~42.7 & -0.334 & M3.7 & 0.9 \\
 26 & 594 & 05~35~35.61 & -05~29~36.5 & -0.492 & M4.4 & 1.0 \\
 27 & 549 & 05~35~35.14 & -05~21~23.6 & -0.115 & M2.3 & 1.5 \\
 28 & 1271 & 05~35~35.02 & -05~41~01.4 & -0.307 & M3.6 & 0.9 \\
 29 & 1322 & 05~35~34.64 & -05~27~15.0 & -0.692 & M4.7 & 1.0 \\
 30 & 1038 & 05~35~34.61 & -05~15~52.7 & -0.442 & M4.3 & 0.9 \\
 31 & 879 & 05~35~34.37 & -05~26~59.6 & -0.814 & M4.9 & 0.9 \\
 32 & 293 & 05~35~34.21 & -05~27~18.2 & -0.334 & M3.7 & 0.9 \\
 33 & 947 & 05~35~33.59 & -05~15~23.2 & -0.045 & M1.8 & 2.0 \\
 34 & 734 & 05~35~33.41 & -05~39~21.3 & -0.186 & M2.8 & 1.2 \\
 35 & 545 & 05~35~33.31 & -05~39~24.9 & -0.086 & M2.1 & 1.7 \\
 36 & 664 & 05~35~33.11 & -05~17~34.0 & -0.525 & M4.4 & 1.0 \\
 37 & 1313 & 05~35~32.88 & -05~39~19.1 & -0.748 & M4.7 & 1.1 \\
 38 & 1459 & 05~35~32.60 & -05~40~12.4 & -0.684 & M4.7 & 1.0 \\
 39 & 1116 & 05~35~31.75 & -05~16~39.9 & -0.038 & M1.6 & 2.1 \\
 40 & 838 & 05~35~31.68 & -05~42~46.0 & -0.239 & M3.2 & 1.1 \\
 41 & 814 & 05~35~31.56 & -05~16~36.9 & -0.978 & M5.6 & 0.7 \\
 42 & 1108 & 05~35~31.53 & -05~15~23.6 & -0.140 & M2.5 & 1.4 \\
 43 & 657 & 05~35~30.70 & -05~18~07.1 & -0.050 & M1.9 & 1.9 \\
 44 & 1011 & 05~35~30.63 & -05~15~16.3 & -0.789 & M4.8 & 1.0 \\
 45 & 460 & 05~35~30.42 & -05~34~38.6 & -0.564 & M4.5 & 1.0 \\
 46 & 364 & 05~35~29.80 & -05~16~06.4 & -0.038 & M1.6 & 2.1 \\
 47 & 1517 & 05~35~29.67 & -05~30~24.7 & -0.301 & M3.6 & 0.9 \\
 48 & 794 & 05~35~28.77 & -05~41~34.0 & -0.230 & M3.1 & 1.1 \\
 49 & 1009 & 05~35~28.28 & -05~37~19.6 & -0.644 & M4.7 & 1.0 \\
 50 & 487 & 05~35~27.66 & -05~42~55.2 & -0.307 & M3.6 & 0.9 \\
 51 & 432 & 05~35~27.20 & -05~30~24.7 & -0.518 & M4.4 & 1.0 \\
 52 & 638 & 05~35~26.72 & -05~16~45.1 & -0.405 & M4.1 & 0.9 \\
 53 & 780 & 05~35~26.45 & -05~30~16.4 & -0.358 & M3.9 & 0.9 \\
 54 & 939 & 05~35~25.53 & -05~34~04.7 & -0.067 & M2.0 & 1.8 \\
 55 & 823 & 05~35~25.33 & -05~25~29.4 & -0.354 & M3.8 & 0.9 \\
 56 & 725 & 05~35~25.22 & -05~29~51.6 & -0.461 & M4.3 & 0.9 \\
 57 & 284 & 05~35~24.64 & -05~11~58.3 & -0.253 & M3.3 & 1.0 \\
 58 & 484 & 05~35~24.47 & -05~11~58.0 & -0.469 & M4.3 & 0.9 \\
 59 & 644 & 05~35~24.09 & -05~21~32.7 & -0.499 & M4.4 & 1.0 \\
 60 & 948 & 05~35~23.50 & -05~34~23.4 & -0.435 & M4.3 & 0.9 \\
 61 & 1054 & 05~35~22.68 & -05~16~14.0 & -0.072 & M2.0 & 1.8 \\
 62 & 341 & 05~35~22.61 & -05~14~11.2 & -0.143 & M2.5 & 1.4 \\
 63 & 1015 & 05~35~22.24 & -05~18~08.8 & -0.735 & M4.7 & 1.0 \\
 64 & 1337 & 05~35~22.07 & -05~28~55.6 & -0.341 & M3.8 & 0.9 \\
 65 & 667 & 05~35~21.97 & -05~17~04.9 & -0.560 & M4.5 & 1.0 \\
 66 & 743 & 05~35~21.88 & -05~17~03.4 & -0.428 & M4.2 & 0.9 \\
 67 & 1347 & 05~35~21.63 & -05~17~19.1 & -0.288 & M3.5 & 1.0 \\
 68 & 713 & 05~35~21.62 & -05~26~57.5 & -0.474 & M4.4 & 0.9 \\
 69 & 560 & 05~35~21.62 & -05~34~58.4 & -0.333 & M3.7 & 0.9 \\
 70 & 420 & 05~35~21.25 & -05~42~12.3 & -0.151 & M2.5 & 1.3 \\
 71 & 795 & 05~35~21.15 & -05~18~21.3 & -0.378 & M4.0 & 0.9 \\
 72 & 415 & 05~35~20.93 & -05~40~14.3 & -0.521 & M4.4 & 1.0 \\
 73 & 739 & 05~35~20.56 & -05~20~43.2 & -0.084 & M2.1 & 1.7 \\
 74 & 1314 & 05~35~19.86 & -05~31~03.7 & -0.178 & M2.8 & 1.3 \\
 75 & 770 & 05~35~19.81 & -05~22~21.6 & -0.084 & M2.1 & 1.7 \\
 76 & 491 & 05~35~19.79 & -05~30~37.5 & -0.563 & M4.5 & 1.0 \\
 77 & 836 & 05~35~19.75 & -05~39~35.6 & -0.235 & M3.2 & 1.1 \\
 78 & 1246 & 05~35~19.56 & -05~27~35.6 & -0.030 & M1.4 & 2.3 \\
 79 & 408 & 05~35~19.48 & -05~36~51.8 & -0.276 & M3.4 & 1.0 \\
 80 & 1486 & 05~35~19.32 & -05~16~09.9 & -0.198 & M2.9 & 1.2 \\
 81 & 414 & 05~35~18.83 & -05~14~45.6 & -0.211 & M3.0 & 1.1 \\
 82 & 1185 & 05~35~18.49 & -05~42~30.7 & -0.516 & M4.4 & 1.0 \\
 83 & 1104 & 05~35~18.46 & -05~39~19.8 & -0.522 & M4.4 & 1.0 \\
 84 & 454 & 05~35~18.29 & -05~28~46.1 & -0.143 & M2.5 & 1.4 \\
 85 & 1070 & 05~35~17.97 & -05~16~45.1 & -0.095 & M2.2 & 1.6 \\
 86 & 523 & 05~35~17.93 & -05~25~33.8 & -0.186 & M2.8 & 1.2 \\
 87 & 666 & 05~35~17.66 & -05~23~41.0 & -1.033 & M5.9 & 0.6 \\
 88 & 433 & 05~35~17.53 & -05~40~48.3 & -0.248 & M3.3 & 1.0 \\
 89 & 552 & 05~35~17.43 & -05~30~25.3 & -0.119 & M2.3 & 1.5 \\
 90 & 468 & 05~35~17.34 & -05~42~14.6 & -0.076 & M2.0 & 1.7 \\
 91 & 707 & 05~35~17.15 & -05~41~53.8 & -0.492 & M4.4 & 1.0 \\
 92 & 723 & 05~35~17.00 & -05~15~44.2 & -0.396 & M4.1 & 0.9 \\
 93 & 950 & 05~35~16.81 & -05~39~17.0 & -0.582 & M4.5 & 1.0 \\
 94 & 68 & 05~35~16.75 & -05~24~04.2 & -0.007 & M0.9 & 2.7 \\
 95 & 686 & 05~35~15.95 & -05~16~57.5 & -0.222 & M3.1 & 1.1 \\
 96 & 709 & 05~35~15.94 & -05~41~11.4 & -0.441 & M4.3 & 0.9 \\
 97 & 1588 & 05~35~15.49 & -05~22~42.9 & -0.048 & M1.8 & 1.9 \\
 98 & 642 & 05~35~15.48 & -05~35~11.9 & -0.157 & M2.6 & 1.3 \\
 99 & 375 & 05~35~15.30 & -05~39~56.1 & -0.329 & M3.7 & 0.9 \\
100 & 936 & 05~35~14.63 & -05~16~46.1 & -0.510 & M4.4 & 1.0 \\
101 & 912 & 05~35~13.75 & -05~34~54.9 & -0.401 & M4.1 & 0.9 \\
102 & 576 & 05~35~13.65 & -05~28~46.2 & -0.189 & M2.8 & 1.2 \\
103 & 693 & 05~35~13.56 & -05~27~57.2 & -0.482 & M4.4 & 0.9 \\
104 & 1130 & 05~35~13.31 & -05~37~15.8 & -0.792 & M4.8 & 1.0 \\
105 & 1274 & 05~35~13.17 & -05~36~18.0 & -0.218 & M3.1 & 1.1 \\
106 & 1078 & 05~35~12.94 & -05~28~49.8 & -0.241 & M3.2 & 1.0 \\
107 & 1648 & 05~35~12.40 & -05~24~03.7 & -0.317 & M3.6 & 0.9 \\
108 & 681 & 05~35~12.26 & -05~20~45.2 & -0.246 & M3.3 & 1.0 \\
109 & 917 & 05~35~12.04 & -05~14~14.6 & -0.179 & M2.8 & 1.3 \\
110 & 662 & 05~35~11.77 & -05~21~55.5 & -0.470 & M4.4 & 0.9 \\
111 & 942 & 05~35~11.72 & -05~23~51.9 & -0.471 & M4.4 & 0.9 \\
112 & 706 & 05~35~11.65 & -05~31~01.1 & -0.485 & M4.4 & 0.9 \\
113 & 1222 & 05~35~11.23 & -05~41~36.1 & -0.191 & M2.8 & 1.2 \\
114 & 721 & 05~35~11.17 & -05~19~35.8 & -0.397 & M4.1 & 0.9 \\
115 & 1168 & 05~35~11.07 & -05~41~56.3 & -0.624 & M4.6 & 1.0 \\
116 & 915 & 05~35~10.90 & -05~22~46.4 & -0.169 & M2.7 & 1.3 \\
117 & 984 & 05~35~10.41 & -05~19~52.4 & -0.186 & M2.8 & 1.2 \\
118 & 482 & 05~35~10.13 & -05~22~32.6 & -0.056 & M1.9 & 1.9 \\
119 & 808 & 05~35~09.27 & -05~16~56.0 & -0.098 & M2.2 & 1.6 \\
120 & 690 & 05~35~08.29 & -05~24~34.9 & -0.613 & M4.6 & 1.0 \\
121 & 827 & 05~35~08.03 & -05~36~14.1 & -0.628 & M4.7 & 1.0 \\
122 & 1272 & 05~35~07.09 & -05~42~33.3 & -0.299 & M3.6 & 0.9 \\
123 & 941 & 05~35~06.91 & -05~26~00.5 & -0.722 & M4.7 & 1.0 \\
124 & 1064 & 05~35~06.83 & -05~42~35.4 & -0.283 & M3.5 & 1.0 \\
125 & 898 & 05~35~06.42 & -05~27~04.7 & -0.741 & M4.7 & 1.1 \\
126 & 1348 & 05~35~05.75 & -05~35~22.1 & -0.739 & M4.7 & 1.0 \\
127 & 413 & 05~35~05.69 & -05~25~04.1 & -0.346 & M3.8 & 0.9 \\
128 & 1123 & 05~35~05.67 & -05~43~04.6 & -0.594 & M4.6 & 1.0 \\
129 & 945 & 05~35~05.61 & -05~18~24.8 & -0.378 & M4.0 & 0.9 \\
130 & 513 & 05~35~05.13 & -05~20~24.4 & -0.130 & M2.4 & 1.5 \\
131 & 1753 & 05~35~03.79 & -05~24~54.4 & -0.435 & M4.3 & 0.9 \\
132 & 1403 & 05~35~02.99 & -05~38~40.5 & -0.613 & M4.6 & 1.0 \\
133 & 1209 & 05~35~02.69 & -05~32~24.9 & -0.597 & M4.6 & 1.0 \\
134 & 1150 & 05~35~01.71 & -05~27~09.8 & -0.802 & M4.9 & 1.0 \\
135 & 1035 & 05~35~01.32 & -05~18~21.3 & -0.436 & M4.3 & 0.9 \\
136 & 503 & 05~35~01.16 & -05~29~55.2 & -0.243 & M3.2 & 1.0 \\
137 & 1540 & 05~34~59.18 & -05~41~12.5 & -0.296 & M3.5 & 0.9 \\
138 & 1572 & 05~34~57.36 & -05~32~43.1 & -0.031 & M1.5 & 2.3 \\
139 & 522 & 05~34~57.23 & -05~42~02.7 & -0.052 & M1.9 & 1.9 \\
140 & 1152 & 05~34~53.48 & -05~40~03.9 & -0.525 & M4.4 & 1.0 \\
141 & 1603 & 05~34~53.07 & -05~26~27.7 & -0.619 & M4.6 & 1.0 \\
142 & 1224 & 05~34~52.67 & -05~21~25.2 & -0.395 & M4.0 & 0.9 \\
143 & 1067 & 05~34~52.36 & -05~25~00.7 & -0.940 & M5.5 & 0.7 \\
144 & 626 & 05~34~52.34 & -05~30~07.9 & -0.384 & M4.0 & 0.9 \\
145 & 331 & 05~34~51.74 & -05~39~24.0 & -0.106 & M2.2 & 1.6 \\
146 & 735 & 05~34~50.86 & -05~39~29.2 & -0.482 & M4.4 & 0.9 \\
147 & 1363 & 05~34~50.21 & -05~35~39.0 & -0.789 & M4.8 & 1.0 \\
148 & 1138 & 05~34~48.45 & -05~31~07.2 & -0.336 & M3.7 & 0.9 \\
149 & 1543 & 05~34~47.73 & -05~26~32.1 & -0.434 & M4.2 & 0.9 \\
150 & 1510 & 05~34~47.66 & -05~31~11.7 & -0.331 & M3.7 & 0.9 \\
151 & 1562 & 05~34~46.87 & -05~30~19.7 & -0.501 & M4.4 & 1.0 \\
152 & 481 & 05~34~45.87 & -05~41~09.6 & -0.212 & M3.0 & 1.1 \\
153 & 307 & 05~34~42.73 & -05~28~37.5 & -0.154 & M2.6 & 1.3 \\
154 & 1366 & 05~34~42.48 & -05~22~46.2 & -0.623 & M4.6 & 1.0 \\
155 & 1242 & 05~34~42.19 & -05~33~03.5 & -0.217 & M3.0 & 1.1 \\
156 & 1243 & 05~34~41.96 & -05~21~32.0 & -0.654 & M4.7 & 1.0 \\
157 & 1160 & 05~34~41.72 & -05~36~48.7 & -0.509 & M4.4 & 1.0 \\
158 & 1211 & 05~34~39.11 & -05~34~02.3 & -0.739 & M4.7 & 1.0 \\
159 & 969 & 05~34~38.19 & -05~35~49.5 & -0.437 & M4.3 & 0.9 \\
160 & 181 & 05~34~36.54 & -05~36~17.1 & -0.008 & M1.0 & 2.7 \\
161 & 661 & 05~34~35.76 & -05~40~09.4 & -0.155 & M2.6 & 1.3 \\
162 & 702 & 05~34~35.68 & -05~35~52.1 & -0.180 & M2.8 & 1.3 \\
163 & 1230 & 05~34~35.60 & -05~37~24.0 & -0.597 & M4.6 & 1.0 \\
164 & 655 & 05~34~33.72 & -05~40~22.7 & -0.595 & M4.6 & 1.0 \\
165 & 858 & 05~34~32.23 & -05~41~48.4 & -0.400 & M4.1 & 0.9 \\
166 & 1155 & 05~34~30.24 & -05~17~01.2 & -0.550 & M4.4 & 1.0 \\
167 & 1052 & 05~34~30.11 & -05~40~15.2 & -0.618 & M4.6 & 1.0 \\
168 & 979 & 05~34~29.34 & -05~33~39.7 & -0.128 & M2.4 & 1.5 \\
169 & 1121 & 05~34~28.96 & -05~23~48.1 & -0.636 & M4.7 & 1.0 \\
170 & 1391 & 05~34~28.56 & -05~30~32.4 & -0.409 & M4.1 & 0.9 \\
171 & 570 & 05~34~27.80 & -05~42~10.2 & -0.025 & M1.3 & 2.4 \\
172 & 401 & 05~34~27.53 & -05~28~28.4 & -0.425 & M4.2 & 0.9 \\
173 & 565 & 05~34~26.75 & -05~41~57.3 & -0.194 & M2.9 & 1.2 \\
174 & 995 & 05~34~25.78 & -05~35~46.5 & -0.411 & M4.1 & 0.9 \\
175 & 949 & 05~34~25.54 & -05~37~02.3 & -0.350 & M3.8 & 0.9 \\
176 & 907 & 05~34~24.05 & -05~42~22.1 & -0.295 & M3.5 & 0.9 \\
177 & 688 & 05~34~21.23 & -05~35~34.7 & -0.502 & M4.4 & 1.0 \\
178 & 902 & 05~34~18.68 & -05~37~08.1 & -0.451 & M4.3 & 0.9 \\
179 & 1708 & 05~34~18.37 & -05~22~54.9 & -0.297 & M3.5 & 0.9 \\
180 & 256 & 05~34~17.48 & -05~31~58.6 & -0.138 & M2.5 & 1.4 \\
181 & 525 & 05~34~17.43 & -05~30~35.2 & -0.189 & M2.8 & 1.2 \\
182 & 1418 & 05~34~17.26 & -05~22~36.8 & -0.022 & M1.3 & 2.4 \\
183 & 1489 & 05~34~17.15 & -05~37~11.8 & -0.269 & M3.4 & 1.0 \\
184 & 388 & 05~34~17.14 & -05~38~16.8 & -0.237 & M3.2 & 1.1 \\
185 & 1053 & 05~34~16.95 & -05~30~53.2 & -0.811 & M4.9 & 0.9 \\
186 & 1294 & 05~34~15.09 & -05~23~00.0 & -0.309 & M3.6 & 0.9 \\
187 & 615 & 05~34~13.87 & -05~36~35.3 & -0.479 & M4.4 & 0.9 \\
188 & 1017 & 05~34~13.51 & -05~35~38.6 & -0.050 & M1.9 & 1.9 \\
189 & 395 & 05~34~13.20 & -05~33~53.5 & -0.131 & M2.4 & 1.5 \\
190 & 889 & 05~34~12.71 & -05~41~36.4 & -0.446 & M4.3 & 0.9 \\
191 & 1086 & 05~34~12.33 & -05~41~34.7 & -0.489 & M4.4 & 0.9 \\
192 & 864 & 05~34~12.02 & -05~24~19.6 & -0.284 & M3.5 & 1.0 \\
193 & 1113 & 05~34~11.50 & -05~30~19.8 & -0.784 & M4.8 & 1.0 \\
194 & 457 & 05~34~09.01 & -05~24~05.6 & -0.065 & M2.0 & 1.8 \\
195 & 121 & 05~34~08.22 & -05~11~43.0 & -0.168 & M2.7 & 1.3 \\
196 & 641 & 05~34~07.96 & -05~36~17.0 & -0.288 & M3.5 & 1.0 \\
197 & 839 & 05~34~07.80 & -05~22~32.3 & -0.216 & M3.0 & 1.1 \\
198 & 358 & 05~34~07.46 & -05~13~36.4 & -0.139 & M2.5 & 1.4 \\
199 & 1099 & 05~34~07.22 & -05~29~32.4 & -0.636 & M4.7 & 1.0 \\
200 & 1032 & 05~34~07.13 & -05~22~27.0 & -0.778 & M4.8 & 1.0 \\
201 & 1073 & 05~34~07.12 & -05~15~59.3 & -0.357 & M3.9 & 0.9 \\
202 & 1457 & 05~34~07.02 & -05~30~20.0 & -0.322 & M3.7 & 0.9 \\
203 & 1090 & 05~34~06.86 & -05~23~08.3 & -0.808 & M4.9 & 0.9 \\
204 & 1508 & 05~34~06.78 & -05~21~46.8 & -0.472 & M4.4 & 0.9 \\
205 & 1057 & 05~34~04.64 & -05~22~22.4 & -0.357 & M3.9 & 0.9 \\
206 & 918 & 05~34~04.40 & -05~36~26.4 & -0.650 & M4.7 & 1.0 \\
207 & 1257 & 05~34~03.89 & -05~29~51.1 & -0.618 & M4.6 & 1.0 \\
208 & 480 & 05~34~03.71 & -05~22~18.7 & -0.236 & M3.2 & 1.1 \\
209 & 935 & 05~34~02.91 & -05~39~21.1 & -0.456 & M4.3 & 0.9 \\
210 & 1343 & 05~34~02.67 & -05~33~02.9 & -0.854 & M5.1 & 0.9 \\
211 & 647 & 05~33~57.91 & -05~36~26.9 & -0.284 & M3.5 & 1.0 \\
212 & 362 & 05~33~57.68 & -05~40~06.0 & -0.098 & M2.2 & 1.6 \\
213 & 427 & 05~33~56.77 & -05~21~33.4 & -0.410 & M4.1 & 0.9 \\
214 & 946 & 05~33~56.56 & -05~39~04.3 & -0.389 & M4.0 & 0.9 \\
215 & 1241 & 05~33~53.97 & -05~27~34.5 & -0.083 & M2.1 & 1.7 \\
216 & 553 & 05~33~53.08 & -05~35~15.7 & -0.471 & M4.4 & 0.9 \\
217 & 384 & 05~33~52.11 & -05~30~28.3 & -0.268 & M3.4 & 1.0 \\
\enddata
\tablenotetext{(*)}{IDs (second column) are as in Table \ref{table:catalog}.}
\label{table:tiostars}
\end{longtable}

\subsection{The H$\alpha$ photometry\\ }
\label{section:halpha}
T Tauri stars are known to exhibit strong H$\alpha$ emission, associated, together with UV excess, with ongoing mass accretion from a circumstellar disk to the star. The H$\alpha$ excess is the most used observational quantity to estimate the mass accretion rates, an important parameter for understanding the evolution of both stars and disks. Here we describe the derived line excess derived from our photometry in the WFI H$\alpha$ filter.

After having calibrated the H$\alpha$ photometry into units of Jy, as described in \S\ref{section:narrowbandcalibration}, we derive the line excess, following an approach analogous to the one used for the TiO index. We computed the flux at 6563 \AA\ by linear interpolation between the (logarithmic) fluxes in $V$ and $I$.  We computed for all stars the ratio between the measured flux and the interpolated flux; for stars that do not emit in H$\alpha$ -- or for which the line emission is negligible -- this ratio should be close to unity. We analyzed the behavior of our assumption of linear interpolation for the photospheric flux by plotting the H$\alpha$ excess as a function of TiO index, to highlight eventual trends with spectral class. In other words, it is reasonable to question the hypothesis that a relation of type (V-I) vs (6563$\AA$-I) is linear along the entire range in colors, especially for late type stars.

\begin{figure}
\epsscale{0.9}
\plotone{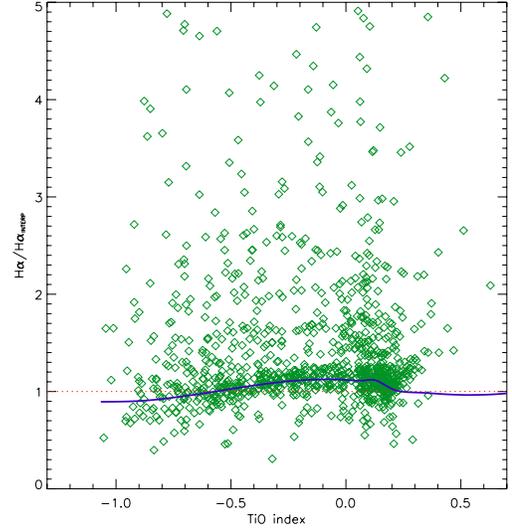}
\caption{Ratio between the measured flux in H$\alpha$ and the one interpolated between $V$ and $I$, with respect to the [TiO] index. Although it is almost equal to unity for a significant fraction of stars that do not show excess (which means that the $V-I$ interpolation at 6563 \AA\ is a good approximation for the intrinsic photospheric flux at this wavelength), the TiO index -- and its correlation with spectral type -- can be used to refine the level of intrinsic stellar flux. The thick solid line, obtained iterating a sigma-clipping robust mean computation technique, represents the zero-emission level at H$\alpha$ as a function of TiO index. \label{fig:excessm_vs_excessh_new}}
\end{figure}
\begin{figure}
\plotone{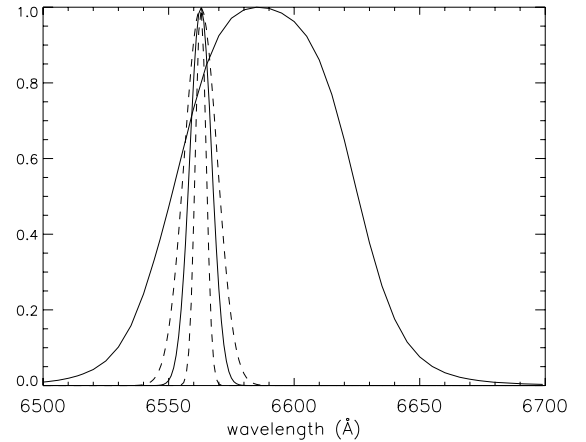}
\caption{The WFI H$\alpha$ filter profile (broad solid line) is compared to H$\alpha$ emission lines, assumed gaussian for simplicity, of different widths: 200km/s (thick solid line), 100km/s and 300km/s (dashed lines). The filter central wavelength is located about 25$\AA$ to the red with respect to $\lambda_{\textrm{H}\alpha}=6563\AA$, where the transparency of the filter is about 80\% of its peak. For a typical value of broadening for the H$\alpha$ emission of CTTS of $\Delta \lambda\lesssim$ 200km/s the , and therefore this average value of filter transparency can be reasonably considered independent of the line broadening. \label{fig:hafilter}}
\end{figure}
\begin{figure*}
\plotone{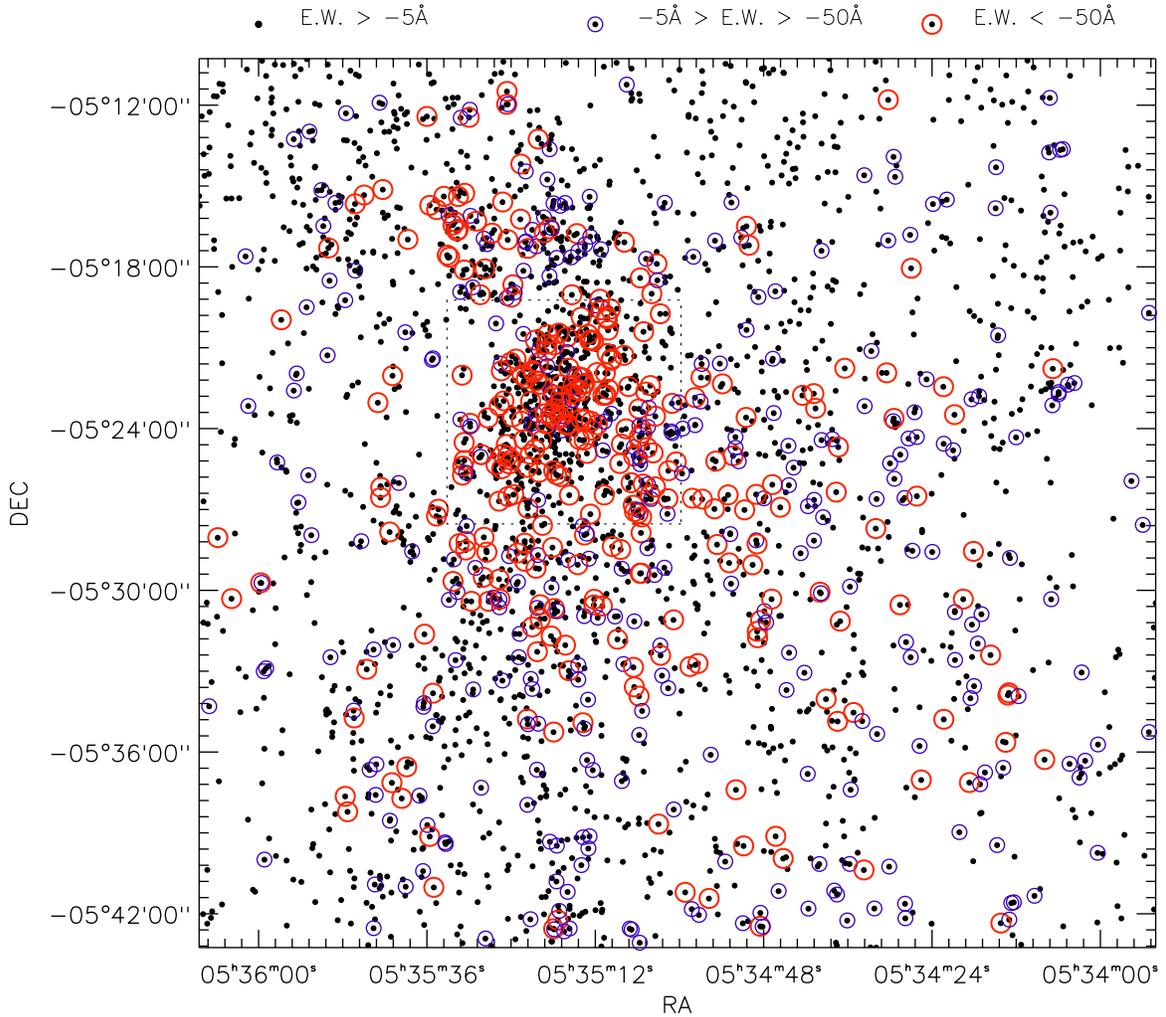}
\caption{Distribution of the stars associated to a H$\alpha$ excess. Black points have E.W. $>$ -5$\AA$ and represent absorption and very weak emission sources; 315 stars are characterized by a weak excess - that we defined to be such that the measured equivalent width of the line emission (in absolute value) is between 5$\AA\ $ and 50$\AA\ $ - (small blue circles), while 323 stars show a strong line excess, exceeding 50$\AA$ in |E.W.| (large red circles).   \label{fig:plotradecHa}}
\end{figure*}

\begin{figure*}
\plotone{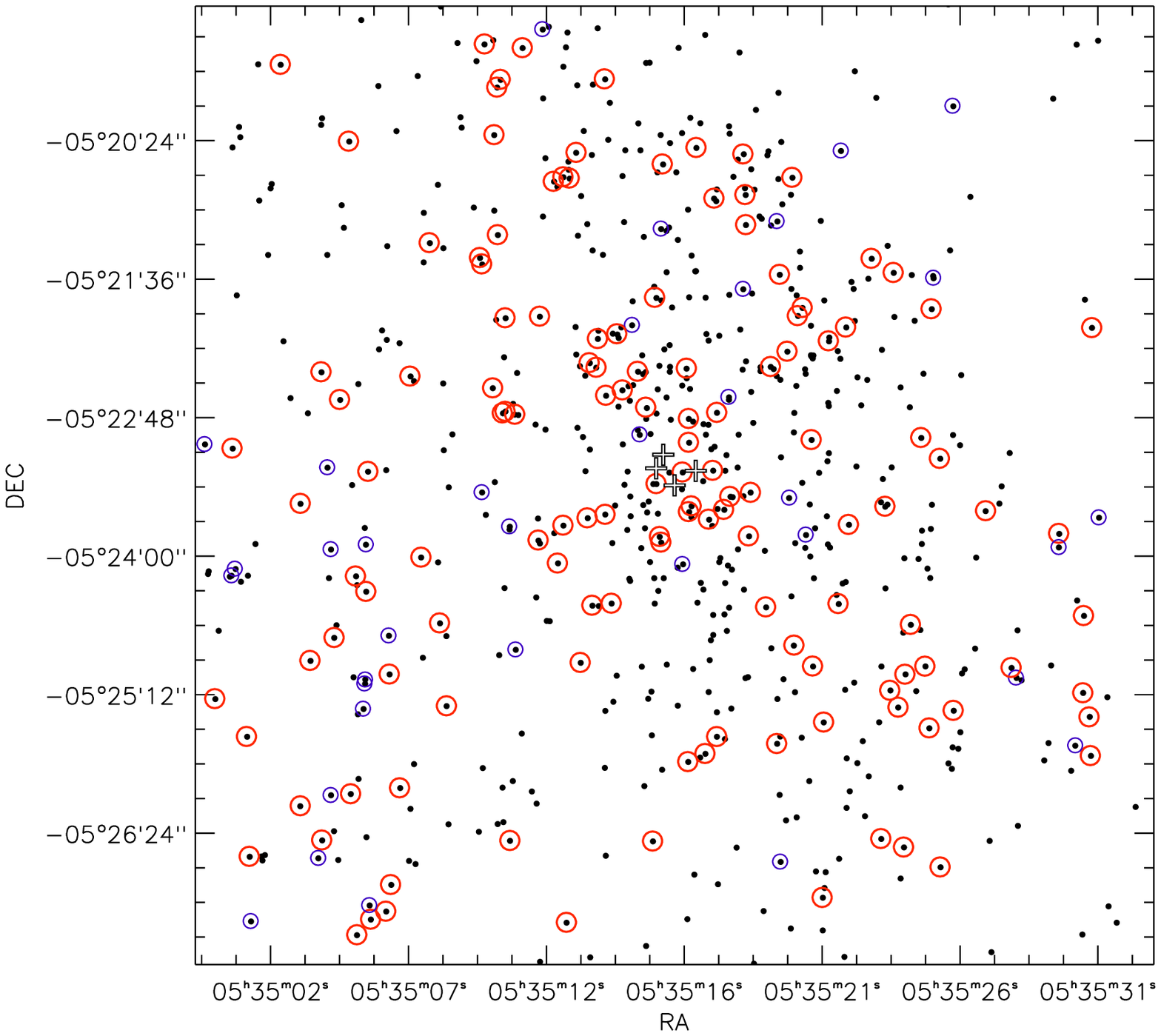}
\caption{Same as Figure \ref{fig:plotradecHaCORE}, for the inner (1pc x 1pc) part of the ONC, centered on the trapezium cluster (denoted by crosses).  \label{fig:plotradecHaCORE}}
\end{figure*}
Indeed, as seen in Figure \ref{fig:excessm_vs_excessh_new}, besides the presence of a large population of stars with a noticeable excess, all the stars without emission lie on a locus which has a weak -- but definite -- dependence on the TiO index, and therefore on spectral type (see [TiO] - temperature relation described in Section \ref{section:TiO}). We isolated this curve iterating a sigma-clipping algorithm and considered this flux level as the photospheric baseline. Clearly this baseline is valid on a statistical base, given that the real photospheric flux in proximity of the H$\alpha$ can have a small dependence on other stellar parameters besides the V and I magnitudes and the [TiO]-derived spectral type, but our estimate remains the best guess relying on our own data.

The line excess, in mJy, does not yet represent the true line emission of the stars, because the filter profile is not centered at $\lambda_{{\rm H}\alpha}$ but is displaced $~25$ \AA\ towards redder wavelengths, and at the line wavelength the filter throughput is about $~80\%$ of its peak. While for the photospheric baseline, locally constant, one can approximate the filter throughput as a squared profile, neglecting the true shape of the filter, this local variation of efficiency must be taken into account, increasing the flux excess by a factor of approximately $20\%$. We investigated the possibility that, in case of a very broad H$\alpha$ line in emission, part of the flux could end up in a part of the filter profile characterized by a very low transmission, limiting the accuracy of the correction factor we applied to the flux. However, CTTSs tend to have H$\alpha$ line emissions with a relatively limited range of broadenings, usually not exceeding $\sim 200$km/s ($4$--$5\AA$), \citep[e.g.][]{alencar2000} much lower than the WFI H$\alpha$ filter width. In Figure \ref{fig:hafilter} we show the filter throughput as a function of wavelength with three examples of $H\alpha$ lines (assumed gaussian for simplicity) overplotted, with a sigma of $100$, $200$ and $300$km/s. It is evident that, even for a relatively high line broadening, the fraction of line flux that ends up in a region of the filter characterized by a very low transparency is negligible, and that, within the line, the throughput can be considered quite linear. Therefore our assumption that the overall transparency of the filter for a H$\alpha$ line in emission can be considered constant and independent on the line broadening is fairly good, or, at least, not the major source of uncertainty in this result.

We correct for this factor and present our the result in Table \ref{table:haexcess}, reporting the excesses both in physical flux and in equivalent width (E.W.)\footnote{For the equivalent width we follow the standard convention, i.e. positive E.W. is associated with an absorption, negative E.W. with an emission.}, as well as the estimated photospheric continuum level.

\begin{deluxetable*}{rrrrrrrrrr}
\tablecaption{H$\alpha$ excess}
\tablehead{
\multicolumn{1}{c}{} &
\multicolumn{1}{c}{ID} &
\multicolumn{1}{c}{RA} &
\multicolumn{1}{c}{Dec} &
\multicolumn{2}{c}{E.W.} &
\multicolumn{2}{c}{H$\alpha$ excess} &
\multicolumn{2}{c}{H$\alpha$ cont.} \\
\multicolumn{1}{c}{} &
\multicolumn{1}{c}{} &
\multicolumn{1}{c}{(J2000)} &
\multicolumn{1}{c}{(J2000)} &
\multicolumn{1}{c}{($\AA$)} &
\multicolumn{1}{c}{(Hz)} &
\multicolumn{1}{c}{(erg/s/cm$^2$)} &
\multicolumn{1}{c}{mJy} &
\multicolumn{1}{c}{(erg/s/cm$^2$)} &
\multicolumn{1}{c}{mJy}}
\startdata
 1 & 911 & 05~36~07.04 & -05~34~18.3 & -49.3 & -3.43E+012 & 1.69E-014 & 0.317 & 2.63E-014 & 0.493 \\
 2 & 1219 & 05~36~05.82 & -05~28~02.9 & -114. & -7.94E+012 & 1.55E-014 & 0.290 & 1.04E-014 & 0.195 \\
 3 & 1136 & 05~36~05.75 & -05~18~55.9 & -1.84 & -1.28E+011 & 3.10E-016 & 0.00582 & 1.29E-014 & 0.242 \\
 4 & 622 & 05~36~05.19 & -05~21~34.5 & 6.83 & 4.75E+011 & -2.12E-015 & -0.0398 & 2.38E-014 & 0.446 \\
 5 & 832 & 05~36~03.85 & -05~30~18.7 & -108. & -7.50E+012 & 5.05E-014 & 0.948 & 3.59E-014 & 0.673 \\
 6 & 1298 & 05~36~03.47 & -05~24~35.9 & 19.6 & 1.37E+012 & -1.34E-015 & -0.0251 & 5.22E-015 & 0.0980 \\
 7 & 909 & 05~36~03.36 & -05~24~23.2 & 3.50 & 2.44E+011 & -1.10E-015 & -0.0206 & 2.40E-014 & 0.450 \\
 8 & 1111 & 05~36~02.82 & -05~32~54.3 & 6.15 & 4.28E+011 & -7.32E-016 & -0.0137 & 9.12E-015 & 0.171 \\
 9 & 887 & 05~36~02.75 & -05~15~27.1 & -1.12 & -7.79E+010 & 2.00E-016 & 0.00375 & 1.37E-014 & 0.256 \\
 10 & 1581 & 05~36~02.59 & -05~34~31.3 & 9.48 & 6.60E+011 & -4.59E-016 & -0.00861 & 3.70E-015 & 0.0695 \\
\nodata & \nodata & \nodata & \nodata & \nodata & \nodata & \nodata & \nodata & \nodata & \nodata \\
\enddata
\tablecomments{Full table available as on-line data. IDs (second column) are as in Table \ref{table:catalog}.}
\label{table:haexcess}
\end{deluxetable*}

In Figures \ref{fig:plotradecHa} and \ref{fig:plotradecHaCORE} we present the sky distribution of the sources for which we detect an excess in the H$\alpha$ flux. A remarkable number of these latter is characterized by a line excess with an equivalent width of more than 50$\AA\ $ in modulus.

Unfortunately an accurate evaluation of the uncertainty in the derived H$\alpha$ emission is limited. Presumably a fraction of the stars can suffer from strong nebular H$\alpha$ contamination, due to the non uniformity of the latter and the relatively low spatial resolution of the seeing limited ground-based observations. Furthermore, emission from unresolved photoevaporated circumstellar disks can lead to an overestimate of the H$\alpha$ excess. A qualitative estimate of at least the first bias is possible from the observed scatter in the ratio between the observed flux and the interpolated continuum level for the stars that don't show line excess. For this purpose we considered all the stars with a positive EW (i.e. H$\alpha$ absorption; these are the points of Figure \ref{fig:excessm_vs_excessh_new} located below the locus where non-emitters should lie (thick line). In an ideal case they should be characterized by a EW$\simeq 0$, however, their scatter, which presents a standard deviation of 12\% and a distribution with tails extended to 50\% of the predicted continuum level, can be considered as representative of the overall uncertainty that we associate to the measured excess. A better estimate of the latter will be possible by means of a cross matching with the photometry obtained with the HST/ACS catalog, for all the sources in the common area, which we postpone to a future publication of our group.

These results will be used, together with the ultraviolet excess obtainable from our WFI U-band photometry, to study the mass accretion rates.


\section{Summary}
\label{section:conclusion}
In this work we present a set of ground-based, simultaneous, broad-band observations of the Orion Nebula Cluster. We produce a catalog in the $U$ $B$, $V$, $I$,  TiO ($\sim 6200\AA$) and H$\alpha$ bands with the Wide Field Imager at the 2.2 m telescope at LaSilla. We publish the calibrated photometry in the WFI instrumental photometric system, corrected for zero-point according to the VegaMag standard for the first 5 bands. The choice of not converting our data into standard systems (such as, for instance, the Johnson-Cousins) is required by the significant diversity of the WFI filter bandpasses, which would require high color term corrections, limiting the accuracy of the transformation because of the dependance of these transformations on stellar parameters. As we are going to show in the second part of this work, this does not affect by any means the practicality of a thorough analysis of our photometric dataset in order to study the PMS population of the ONC.

We define a spectro-photometric index, that we name [TiO] index, from the fluxes in V-, I- and TiO 6200$\AA\ $, and, comparing it with the spectral atlas of ONC members of \citet{hillenbrand97}, we find and present a correlation between the two quantities in the M-type stars range. This allows us to classify 217 new stars, whose spectral type is now presented.

We present the H$\alpha$ photometry, calibrated in terms of energy flux, and, evaluating in a rigorous way the intrinsic photospheric contribution in the luminosity at the  H$\alpha$ wavelength, we derive the line excess. We present therefore the H$\alpha$ excess in terms of Equivalent Width or physical flux for 1040 stars of our catalog, and we discuss the accuracy of the reported values. These latter will be used to derive and study the mass accretion rates in the Orion Nebula Cluster.\\

\acknowledgements
We wish to thank ESO's General Director for awarding DDT\ time to this project and the ESO\ staff in La Silla for carrying out the observations in Service Observing mode.  We  acknowledge the European Southern Observatory for the use of the software {\em alambic} (ESO/MVM).

This work was made possible in part by GO\ program 10246 of the {\it Hubble Space Telescope}, which is operated by the Space Telescope Science Institute.

This work was made possible through the Summer Student Program of the Space Telescope Science Institute.

{\it Facilities:} \facility{ESO}



\end{document}